\documentclass[aps,prl,floatfix,twocolumn,nofootinbib]{revtex4}
\usepackage{dcolumn}
\usepackage{bm}
\usepackage[dvips]{epsfig}
\usepackage{graphicx} \usepackage{amsmath} \usepackage{amssymb}
\newcommand{\comment}[1]{}
\newcommand{\BEQ}{\begin{equation}}
\newcommand{\EEQ}{\end{equation}}
\newcommand{\BEA}{\begin{eqnarray}}
\newcommand{\EEA}{\end{eqnarray}}
\renewcommand{\d}{{\rm d}}

\newcommand{\p}{\bar{p}}
\renewcommand{\v}{\bar{v}}
\renewcommand{\r}{\bar{\rho}}

\renewcommand{\a}{\bar{a}}
\newcommand{\w}{\bar{w}}
\newcommand{\ma}{{\rm M}^2_1}
\renewcommand{\w}{\bar{w}}
\newcommand{\f}{\bar{F}}

\newcommand{\g}{\gamma}
\renewcommand{\gg}{\frac{\gamma}{\gamma-1}}
\newcommand{\vnab}{\vec{\nabla}}

\begin{document}

\title{Work-extraction from fluid flow: the analogue of Carnot's efficiency}

\author{A.E. Allahverdyan} 

\affiliation{Alikhanyan National Laboratory (Yerevan Physics
  Institute), Alikhanian Brothers Street 2, Yerevan 375036, Armenia}

\begin{abstract} 

Aiming to explore physical limits of wind turbines, we develop a model
for determining the work extractable from a compressible fluid
flow. The model employs conservation of mass, energy and entropy and
leads to a universal bound for the efficiency of the work extractable
from kinetic energy. The bound is reached for a sufficiently slow,
weakly-forced quasi-one-dimensional, dissipationless flow. In several respects the
bound is similar to the Carnot limit for the efficiency of heat-engines.
More generally, we show that the maximum work-extraction demands a
contribution from the enthalpy, and is reached for sonic output
velocities and strong forcing.

\end{abstract}


\comment{ 
   }

\maketitle

How much work can be extracted from the kinetic energy of a fluid flow?
The question is old \cite{lanchester_o,betz_o,jouk_o}, but it is still of
obvious practical importance for wind energy usage \cite{handbook}; 
e.g. it is relevant for shaping renewable energy policies \cite{mackay}. 
Wind turbines cannot extract the whole kinetic energy, otherwise the flow will stall.
The question is of fundamental importance, since it asks about the operational 
meaning of energy stored in a continuous medium.

\comment{(for a point particle the whole kinetic energy can be extracted).}

No satisfactory answer to the above question is known. A popular model
developed by Betz \cite{betz_o} (and independently by Lanchester
\cite{lanchester_o} and Joukowsky \cite{jouk_o}) studies a quantity
$\zeta_{\rm B}$, which is smaller than the efficiency of work extracted
from kinetic energy and proposes for it an upper bound $\zeta_{\rm
B}\leq\frac{16}{27}$; see \cite{handbook,mackay,pelka,inglis,sorensen}
for reviews.  Betz's model makes an unwarranted assumption about the
pressure distribution \cite{greet,rauh}. The proper efficiency in the
model is bounded by $1$; see \S 1 and \S 2 of \cite{sup}.  Hence Betz's
model does not answer the question. 

We study work-extraction due to an external force, employing
conservation laws of mass, entropy and energy for a dissipationless, stationary
fluid. The flow model is realistic, since the force is
general, no incompressibility is assumed {\it etc}. Our main assumption
is that the axial component of the flow velocity is homogeneous along
both initial and final cross-sections of the flow. 

We derive a new upper bound for the efficiency of work-extraction from
the kinetic energy. We focus on this form of work-extraction, because it
is relevant for wind turbines
\cite{lanchester_o,betz_o,jouk_o,mackay,handbook,pelka,inglis,sorensen,greet,rauh},
and also because it is similar to heat-engine physics.
The bound is attained for a weakly forced, subsonic, quasi-1d flow,
where the fluid undergoes a cyclic process: its density and pressure
after action of the force are equal to their initial values.  This
resembles Carnot's bound for heat engines that is also reached for
cyclic, slow and dissipationless processes \cite{callen}. We also
determine the maximal work extracted from flow, without demanding that
it necessarily comes from the kinetic energy. The maximum is reached for
sonic output velocities and strong forcing. 
In this regime the work comes from enthalpy and can relate to increasing
kinetic energy.

\begin{figure}[ht]
\includegraphics[width=7cm]{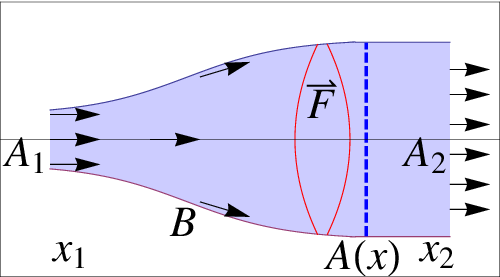}
\caption{
The model. The flow (denoted by blue) goes from $x_{1}$ (input) to
$x_{2}$ (output). $\vec{F}$ is the external force. The control volume
$B$ is filled.  $A(x)$ (dashed line) is the cross-section.  $A_1=A(x_1)$
and $A_2=A(x_2)$ are (resp.) input and output surfaces.  Red curves
bound the domain $\Omega$, where $\vec{F}$ is localized.  Arrows denote
stationary flow velocities. 
}
\label{fig00}
\end{figure}

{\it The model.} The filled domain in Fig.~\ref{fig00} shows the
stationary flow model. Here are our assumptions about it.  

{\bf 1.} The fluid is dissipationless and compressible. 

{\bf 2.} The work-extracting part of the turbine is modeled by a
stationary space-dependent force $\vec{F}(\vec{x})$, which is zero out
of a finite domain $\Omega$; see Fig.~\ref{fig00}. 

{\bf 3.} Homogeneous input flow: at the input $\vec{r}_1\equiv(x_{1},y,z)$, which is
far from $\Omega$ (to the left in Fig.~\ref{fig00}) the pressure $p$, velocity $\vec{v}$ and density
$\rho$  do not depend on $(y,z)$, and transverse velocities are absent:
\begin{gather}
  \vec{v}(\vec{r}_1)=(v_{1},0,0), \qquad p(\vec{r}_1)=p_{1}, \qquad   \rho(\vec{r}_1)=\rho_{1}.
\label{eq:48}
\end{gather}

The control volume $B$ in Fig.~\ref{fig00} is defined along the flow
lines via 2 conditions: {\it (i)} it can be used to calculate the total
work (volume integral) $\int_\Omega \d V \,
(-\vec{v}\cdotp\vec{F})=\int_B \d V\, (-\vec{v}\cdotp\vec{F})$ done by
$\vec{F}$.  {\it (ii)} The area $a(x_1)$ of the input surface
$A_1=A(x_1)$ is possibly small, as needed for ensuring assumption {\bf
5} below, and for calculating the efficiency; see (\ref{vedma2},
\ref{guppi}) below. Hence $B$ encircles $\Omega$; cf.~Fig.~\ref{fig00}. 

{\bf 4.} The cross-section $A(x)$ of $B$ grows with $x$ from input
$A(x_1)$ to output $A(x_2)$. This assumption is needed for achieving work-extraction. The general 
bounds (\ref{balto}, \ref{ragoro}) on the efficiency of work-extraction demand a weaker condition
$a(x_2)>a(x_1)$, where $a(x)$ is the area of $A(x)$.

{\bf 5.} $v_x$ is constant along the output surface $A_{2}$:
\begin{eqnarray}
  \label{basic1}
  \vec{v}(\vec{r}_2)=(v_2, v_y(\vec{r}_2), v_z(\vec{r}_2)), \qquad \vec{r}_2\equiv(x_{2},y,z).
\end{eqnarray}
At the output $\vec{r}_2$ we apply the following notation
\BEA
  \label{basic2}
p(x_2,y,z)=p_2\,\widetilde{p}(y,z), \quad
\rho(x_2,y,z)=\rho_2\,\widetilde{\rho}(y,z),
\EEA
where $\widetilde{p}(y,z)$ and $\widetilde{\rho}(y,z)$ are defined so as to hold
\begin{gather}
\label{avo}
\langle \widetilde{p} \rangle\equiv
\int_{A_2}\frac{{\rm d}y\,{\rm d}z\,\widetilde{p}(y,z)}{a_2}=1,\qquad \langle \widetilde{\rho} \rangle
=1. 
\end{gather}
Eq.~(\ref{basic1}) is a weak form of the plug-flow assumption done in hydraulics
and quasi-1d motion \cite{tommy,landau}; see \cite{luchini,manneville,noble} for reviews that explore
limits of plug-flows. 

{\bf 6.} The fluid is an ideal gas with constant
heat-capacities $c_V$ and $c_p$. This implies for the 
entropy density $s$ and internal energy density $\varepsilon$ \cite{landau}:
\begin{eqnarray}
  \label{eq:5}
\frac{s}{c_V}= \ln p-\g\,\ln \rho,\qquad
\varepsilon
=\frac{1}{\g-1}\,\frac{p}{\rho}, \qquad \g\equiv \frac{c_p}{c_V}>1,
\end{eqnarray}
where the integration constant in $s$ was fixed as in \cite{landau}. 
For air $\g=1.4$ in agreement with the thermodynamic bound $\gamma>1$ \cite{landau}. 
The local speed of sound reads \cite{landau}
\BEA
\label{sound}
v_{\rm s}^2=\left(\left.{\partial p}/{\partial\rho}\right)\right|_s=\g{p}/{\rho}.
\EEA
The set-up is a generalization of Betz's model
\cite{lanchester_o,betz_o,jouk_o,mackay,handbook,pelka,inglis,sorensen,greet,rauh},
because we do not assume that flow is incompressible, and we do not
restrict $\vec{F}$ to be localized in a thin surface. 
Limitations of the set-up are discussed in \S 3 of \cite{sup}. 

\comment{
Assumptions {\bf 4} and {\bf 5} are consistent with dissipationless
boundary conditions; see Fig.~\ref{fig00}. Assumption {\bf 3}
is restrictive, because wind turbines have blades that move faster than
wind, and do not just exert a stationary force \cite{sorensen}.
Both {\bf 4} and {\bf 5} are limited by the
air turbulence \cite{turbu_1,turbu_2}, but the validity of 
{\bf 3}, {\bf 4} and {\bf 5} can be recovered after time-averaging \cite{landau,pelka}. 
Assumption {\bf 4} is also limited by vorticity of input flow. 
}

{\it Conservation laws} of mass, entropy and energy read for stationary flow
\cite{landau} [$\vnab=(\frac{\partial}{\partial x}, \frac{\partial}{\partial y}, \frac{\partial}{\partial z})$]:
\begin{eqnarray}
  \label{eq:1}
&& \vnab\cdotp (\rho\vec{v})=0, \qquad \vnab\cdotp (\rho\vec{v}s)=0, \\
  \label{eq:2}
&& \vnab\cdotp\left[\frac{\rho \vec{v}^2\vec{v}}{2}+\rho(\varepsilon
+\frac{p}{\rho})\vec{v}\right]  =\vec{v}\cdotp\vec{F}, 
\end{eqnarray}
where $\varepsilon +\frac{p}{\rho}$ is the enthalpy density, and where
the external force $\vec{F}$ enters into stationary Euler's equation as:
\BEA
\label{leonard}
\rho\,{\d \vec{v}}/{\d t}=\rho\,(\vec{v}\cdotp\vnab)\vec{v}=-\vec{\nabla}p+\vec{F}.
\EEA
The momentum conservation is not employed, 
since it is useless without restrictive assumptions; see \S
1 of \cite{sup}. 

We apply (\ref{eq:1}, \ref{eq:2}) to the control domain $B$ in Fig.~\ref{fig00}.
Integrate $\vnab\cdotp (\rho\vec{v})=0$ in (\ref{eq:1}) over the volume
$B$ [cf.~Fig.~\ref{fig00}], and employ 
Gauss theorem to get 3 integrals over the surface of $B$:
$(\int_{A_1}+\int_{A_2}+\int_{{\cal B}})\d \vec{n}\cdotp\vec{v}\rho=0$, where $\d
\vec{n}$ points outward. Boundary conditions for a dissipationless
fluid imply $\d \vec{n}\cdotp\vec{v}|_{{\cal B}}=0$ \cite{landau}. Then employ (\ref{eq:48}--\ref{avo})
in $(\int_{A_1}+\int_{A_2})\d \vec{n}\cdotp\vec{v}\rho$.  Other two
relations in (\ref{eq:1}, \ref{eq:2}) are treated in the same way, also
using (\ref{eq:5}):
\BEA
  \label{eq:01a}
 a_{1}\rho_{1} v_{1} =a_2 \rho_2 v_2, \qquad (p_2/p_{1})=(\rho_2/\rho_{1})^{\g}\, e^{\sigma }, \\
  \label{eq:02a}
 \frac{-\int {\rm d}V\, \vec{v}\cdotp\vec{F}  }{a_{1}\rho_{1} v_{1}}=  \frac{v^2_{1}-v^2_{2}-{v}_{\rm tr}^2}{2}+\gg\left(\frac{p_{1}}{\rho_{1}}
-\frac{p_{2}}{\rho_{2}}\right),
\EEA
where $\int {\rm d}V$ goes over  volume $B$ (colored blue in Fig.~\ref{fig00}), $a_k$ is the area of $A_k$, and
where [cf.~(\ref{basic1}--\ref{avo})]
\BEA
  \label{eq:020a}
&& {v}_{\rm tr}^2\equiv \left\langle \widetilde{\rho}\,[\, v^2_y(\vec{r}_2)+v^2_z(\vec{r}_2)   \,]
\right\rangle, \\
  \label{eq:05a}
  &&  \sigma\equiv   \langle\, \widetilde{\rho}\,
  \ln [\,{\widetilde{\rho}}/{\widetilde{p}}\,] \,\rangle+(\g-1)
  \langle\,\widetilde{\rho}\,\ln\widetilde{\rho}\,\rangle\geq0.
\label{tartar}
\EEA
The LHS of (\ref{eq:02a}) is the extracted work that amounts to the kinetic
energy+enthalpy difference between input and output.  Here ${v}_{\rm
tr}^2$ is the output transverse velocity contribution including vorticity. Both terms in
(\ref{tartar}) are non-negative \cite{zhuk} due to spatial inhomogenuities 
at the final surface $A_2$. Now $\sigma$ corresponds to
an effective entropy production [cf.~(\ref{avo})]. If we include
dissipative effects by introducing in (\ref{eq:1}) a non-zero entropy
production $\vnab\cdotp (\rho\vec{v}s)=s_{\rm prod}$, then above formulas will
hold upon $\sigma\to\sigma+\frac{1 }{c_V\,a_{1}\rho_{1} v_{1}}\int {\rm
d}V\, s_{\rm prod}$. Thus even for a dissipationless fluid, the
inhomogeneity of the output plays the role of an effective entropy
production $\sigma>0$. 

To simplify (\ref{eq:01a}, \ref{eq:02a}), employ dimensionless parameters:
\BEA
\label{defoo}
&&\a_2=\frac{a_2}{a_1}, ~~~ 
\v_2=\frac{v_2}{v_1}, ~~~ 
\p_2=\frac{p_2}{p_1}, ~~~ 
\v_{\rm tr}=\frac{v_{\rm tr}}{v_1}, \\
&&\ma= \frac{ \rho_1v_1^2}{\gamma p_1},\quad 
\w\equiv -\frac{\int {\rm d}V\, \vec{v}\cdotp\vec{F}  }{\frac{1}{2}a_{1}  \rho_{1} v^3_{1}}, 
\label{vedma2}
\EEA
where ${\rm M}_1$ (Mach number) is ratio of the input velocity to the
speed of sound (\ref{sound}) at the input, and $\w$ is the dimensionless work defined
as the ratio of the work to the inflow ${\frac{1}{2}a_{1} \rho_{1}
v^3_{1}}$ of kinetic energy.  Eqs.~(\ref{eq:01a}) lead to $\p_2\,
\v_2^\g \, \a_2^\g =e^\sigma$ to be used together with (\ref{defoo}, \ref{vedma2}) in
(\ref{eq:02a}):
\BEA
\label{vedma}
\w=
1-\v_2^2-\v_{\rm tr}^2+\frac{2}{\ma(\g-1)}(1-e^{\sigma}\a_2^{1-\g}\,\v_2^{1-\g}).
\EEA
Our purpose is to extract work, hence to achieve $\w>0$. 

{\it Work-extraction from kinetic energy.} We demand in
(\ref{vedma}) that the work is extracted from kinetic energy only:
\BEA
\label{feofan}
1=e^{\sigma}\a_2^{1-\g}\,\v_2^{1-\g}. 
\EEA
Due to $\sigma>0$ and $\g>1$, condition (\ref{feofan}) 
can be achieved for $v_2<1$ (smaller kinetic energy) only for $\a_2>1$ [cf.~{\bf 4}].
Using (\ref{feofan}) and $\a_2>1$ we get from (\ref{vedma}, \ref{defoo})
\BEA
\label{chav}
\w&=& 1-\a_2^{-2} \, e^{\frac{2\sigma}{\g-1}} -\v_{\rm tr}^2\\
&\leq& 1-\a_2^{-2}=1-(a_1/a_2)^2,
\label{balto}
\EEA
where in deriving (\ref{balto}) we employed $\sigma\geq 0$, $\g>1$ and
$\v_{\rm tr}^2\geq 0$. Once the work is extracted from the kinetic
energy only, the latter is the resource and then (\ref{chav}) is also
the efficiency, i.e. the result over resource. 
Two hindrances for reaching (\ref{balto}) from (\ref{chav}) are
$\v_{\rm tr}^2> 0$ and $\sigma>0$.
The bound (\ref{balto}) is
attained for $\v_{\rm tr}^2= 0$ (no tangential velocity) and $\sigma=0$
(no effective entropy production). The latter relation means
$\widetilde{p}=\widetilde{\rho}=1$; cf.~(\ref{tartar}, \ref{basic2}).
\S 4 of \cite{sup} shows that (\ref{balto}) holds for non-ideal gases. 

Below we demonstrate that bound (\ref{balto}) is attained for quasi-1d 
motion, where $\sigma=\v_{\rm tr}=0$ and $\widetilde{\rho}=\widetilde{p}=1$ take place naturally. Then as 
(\ref{feofan}, \ref{eq:01a}) show, work-extraction from kinetic energy 
demands cyclicality:
\BEA
\label{mito}
\rho_1=\rho_2, \qquad p_1=p_2.
\EEA
Note that only requiring $\rho_1=\rho_2$ we establish the bound (\ref{balto}) from 
(\ref{eq:01a}) and $\sigma\geq 0$. 
The shape of (\ref{balto}), cyclicality condition (\ref{mito}) and no entropy production $\sigma=0$
(needed for attaining (\ref{balto})) make analogy between (\ref{balto}) and Carnot's bound 
for heat-engines. 

{\it Work maximization over the final velocity.} 
The work (\ref{vedma}) is formally maximized over $\v_2$|for
fixed values of other parameters|via
$\frac{\partial \w}{\partial \v_2}=0$ and $\frac{\partial^2
\w}{\partial \v_2^2}<0$. The second relation holds always, while the
first one produces:
\begin{gather}
\label{maxim}
\v_2=\v_{\rm m}=\left( e^\sigma \a_2^{1-\g}{\rm M}_1^{-2} \right)^{\frac{1}{\g+1}},\\
\w_{\rm m}=\w(\v_{\rm m})=1-\v_{\rm m}^2-\v^2_{\rm tr}+\frac{2(1-\ma\,\v_{\rm m}^2)}{(\g-1)\ma},
\label{maximo}
\end{gather}
The output velocity that corresponds to $\v_{\rm m}$ equals to the speed of sound, 
as seen by starting from (\ref{sound}, \ref{eq:01a}, \ref{defoo}):
\BEA
\frac{v^2_{\rm s}(x_2)}{v_1^2}=\frac{1}{\ma}\,\frac{(p_2/p_1)}{(\rho_2/\rho_1)}
=\frac{ e^\sigma \a_2^{1-\g} \v_{\rm m}^{1-\g}}{\ma}=\v_{\rm m}^2.
\label{durban}
\EEA
and noting that the last equality amounts to (\ref{maxim}). 
The maximal work $\w_{\rm m}$ can be attained, as seen below. 

\comment{  
Within (\ref{maxim}, \ref{maximo}) the work-extraction generally takes
place also from enthalpy, since we can have in (\ref{vedma})
\BEA
\label{uber}
1>e^\sigma\a_2^{1-\g} \v_{\rm m}^{1-\g} =e^{\frac{2\sigma}{1+\g}}\left(\ma\,\a_2^{-2}  \right)^{\frac{\g-1}{\g+1}}
\EEA
e.g. due to a small entropy production $\sigma\to 0$ and subsonic initial velocities: $\ma<1$. }

{\it Work-extraction in quasi-1d flow.} Eqs.~(\ref{eq:01a},
\ref{eq:02a}) are useful for bounding the work, but they cannot
determine it, since $v_2$, $\sigma$, and $v_{\rm tr}$ are unknown. A
more specific and informative approach is needed that allows to address
the attainability of bounds.  Since the flow (shown in Fig.~\ref{fig00})
has a smooth and slowly varying cross-section $A(x)$, we apply the
quasi-1d approach \cite{tommy,landau}. It assumes a {stationary} flow
with the axial flow velocity $\vec{v}=(v,0,0)$, pressure $p$, density
$\rho$, and the external force $\vec{F}=(F,0,0)$ depending {only} on the
axial variable $x$. Hence transverse velocities and effective entropy production
nullify: $v_{\rm tr}=\sigma=0$; cf.~(\ref{eq:020a}, \ref{eq:05a}). I.e.
two hindrances for reaching (\ref{balto}) from (\ref{chav}) are absent for the quasi-1d model.

We use scaled functions of $x$ [cf.~(\ref{defoo})]:
\BEA
\v\equiv \frac{v}{v_1}, ~~\, \r\equiv \frac{\rho}{\rho_1}, ~~\,
\p\equiv \frac{p}{p_1}, ~~\, \a\equiv\frac{a}{a_1},~~\, 
\f\equiv \frac{F}{p_1}.
\label{utok}
\EEA
Conservation laws
of mass and entropy \cite{tommy,landau} are to be taken from volume integrals of
(\ref{eq:1}) [cf.~(\ref{eq:01a}, \ref{utok})]
\BEA
\label{gugu}
\r(x)\a(x)\v(x)=1, \qquad
\p(x)=\r^{\gamma}(x).
\EEA
Eqs.~(\ref{gugu}) go together with the stationary Euler
equation (\ref{leonard}) written with the 1d assumption [cf.~(\ref{eq:48}, \ref{vedma2})]:
\BEA
\rho vv'=-p'+F,\quad {\rm or} \quad
\g\ma\,\r\, \v \, \v' =-\p'+\f,
\label{mora}
\EEA
where $\frac{\d {\cal X}}{\d x}\equiv {\cal X}'$ for any ${\cal X}$.
Eqs.~(\ref{mora}, \ref{gugu}) lead to 
\BEA
\left[\frac{\g\ma\,\v^2}{2}+\frac{\gamma \r^{\gamma-1}}{\gamma-1}  \right]'=\frac{\f}{\r}.
\label{berno}
\EEA
The work will be directly calculated from its definition (\ref{eq:02a}, \ref{vedma2}) by employing (\ref{gugu}, \ref{berno}):
\begin{gather}
\label{khud}
\frac{\w\,\g\ma}{2}=-\frac{\int {\rm d}V\, \vec{v}\cdotp\vec{F}}{a_1p_1v_1}=-
\int_{x_1}^{x_2}\d x\, \a(x) \v(x) \f (x) \\
=-\int_{x_1}^{x_2}\d x\, \frac{\f (x)}{\r (x)}=\int_{x_2}^{x_1}\d x\,
\left[\frac{\g\ma\v^2}{2}+\frac{\gamma \r^{\gamma-1}}{\gamma-1}  \right]'.
\label{udo}
\end{gather}
Eq.~(\ref{udo}) recovers the general formula (\ref{vedma}) with
$\sigma=\v_{\rm tr}=0$, as a consequence of the quasi-1d approach.

\comment{ 
Using (\ref{gugu}), we can write (\ref{mora}) as a differential equation for unknown $\v$ given $\f$:
\BEA
\v'=\frac{\g \,\a^{-\g}\, \v^{-\g}\a'+\a\,\f  }{\g\ma-\g\, \a^{1-\g}\,\v^{-\g-1}}.
\label{drob}
\EEA
}

To understand the physics of this problem, let us 
note that (\ref{gugu}) can be written as, respectively,
\BEA
\label{osh}
\frac{\r'}{\r}+\frac{\v'}{\v}+\frac{\a'}{\a}=0,\qquad
\g\frac{\r'}{\r}= \frac{\p'}{\p}.
\EEA
We take the derivative in (\ref{berno}) and work it out in 2 different ways using (\ref{osh}) and $\p(x)=\r^{\gamma}(x)$:
\BEA
\label{rahmon}
&&\frac{p'}{p}\left(1-\frac{v^2}{v^2_{\rm s}}  \right)=\frac{F}{p}+\frac{\g v^2}{v^2_{\rm s}}\,\frac{a'}{a},\\
&&\frac{v'}{v}\left(\frac{v^2}{v^2_{\rm s}} -1 \right)=\frac{a'}{a}+\frac{F}{\g p},
\label{shavkat}
\EEA
where $v_{\rm s}=v_{\rm s}(x)$ is the speed of sound defined in
(\ref{sound}). In the subsonic case ${v^2}<{v^2_{\rm s}}$ consider
firstly (\ref{rahmon}, \ref{shavkat}) for $F=0$ \cite{landau,tommy}. Now
$a'(x)>0$ implies expected trends: $p'(x)>0$ and $v'(x)<0$.
Eqs.~(\ref{rahmon}, \ref{shavkat}) show that a $F<0$ can reverse those
trends for $a'(x)>0$. This reversing will be seen to be the mechanism of
work-extraction. 

\comment{
\BEA
\label{rahmon}
\frac{p' (1-{\rm M}^2)}{p}=\frac{F}{p}+ \frac{a' \g {\rm M}^2}{a},\,
\frac{v' ({\rm M}^2 -1 )}{v}=\frac{a'}{a}+\frac{F}{\g p},
\label{shavkat}
\EEA
}

\begin{figure}[ht]
\includegraphics[width=7cm]{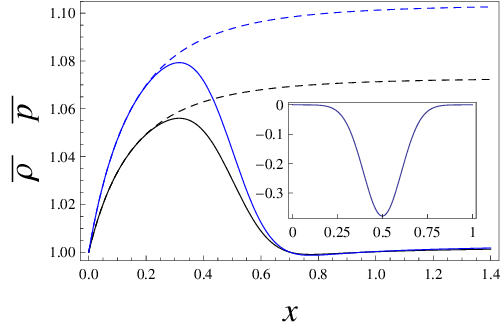}
\caption{ 
Dimensionless density $\r$ (black curve) and pressure $\p$ (blue curve)
versus $x$ obtained from solving (\ref{mora}, \ref{gugu})
with an external force $\f(x)=-\frac{f}{L\sqrt{\pi}} \exp[-(x-x_0)^2/L^2]$.
The force is shown in the inset. Its magnitude is $f=0.1$, center is at
$x_0=0.5$, and $L=0.15$ (we can take $L<0.15$ without serious changes). Other parameters: $x_1=0$,
$\a(x)=(1+x\times 1.5)^2$, $\gamma=1.4$ (air), $\ma=1/7<1$ (subsonic input flow). 
Here $\a(x)=(1+\frac{x}{x_2}\,\frac{r_2-r_1}{r_1})^2$ refers to a truncated-cone shape of $B$ in Fig.~\ref{fig00}
with maximal and minimal radii (resp.) $r_2$ and $r_1$. This is the simplest shape for our ends. 
The black dashed curve and blue dashed curve show (resp.) $\r$ and $\p$
for $\f=0$.\\
The dimensionless velocity $\v(x)$ decays  (not shown) reaching
value $1/\a(x_2)$ for $x=x_2=0.955$; cf.~(\ref{feofan}) with $\sigma=0$. 
The dimensionless work $\w(x)$ (not shown) grows and saturates at 
(\ref{kurdian}) for $x\geq 0.8$.
We choose $x_2=0.955$, since the enthalpy contribution to the work is 
zero. (This contribution also nullifies for $x_2=0.702$, but there the work is smaller.)
The efficiency of work-extraction from kinetic energy equals to its maximal value
(\ref{balto}), which is $0.971$ for the present case. 
}
\label{fig1}
\end{figure} 

Fig.~\ref{fig1} exemplifies the first scenario of work extraction, where
$\f$ is weak.  The velocity $\v(x)$ decays with $x$; its behavior is
close to the case $\f=0$ in (\ref{mora}). But the density $\r(x)$
does feel the weak force, since it changes cyclically returning to the
initial value once the force ceases to act. We define $x_2$ such that
$\r(x_1)=\r(x_2)$; see (\ref{mito}) and Fig.~\ref{fig1}. 
Hence the work is extracted from the kinetic energy only, and the
efficiency equals its maximal value (\ref{balto}). 

Eqs.~(\ref{rahmon}, \ref{shavkat}) explain why the weak force changes
qualitatively the behavior of $\p(x)=\r^{\g}(x)$, but does not change the behavior
of $\v(x)$: the geometric factor $\frac{\a'}{\a^{\,}}$ in (\ref{rahmon})
is multiplied by a factor $\frac{\g v^2}{v^2_{\rm s}}$, which is small
for the subsonic flow, and which is lacking in (\ref{shavkat}). 

Fig.~\ref{fig1} shows that the change of density $\r(x)$ is
small. Hence we can put $\r(x)\simeq 1$ in (\ref{khud}, \ref{gugu}) obtaining 
\BEA
\label{kurdian}  \w\simeq -
\frac{2}{\g\ma}\int_{x_1}^{x_2}\d x\, \f (x). 
\EEA
For parameters of Fig.~\ref{fig1} both work and efficiency 
can be maximized simultaneously. But generally there is a 
conflict between these two maximizations; see
\S 5 of \cite{sup}.

\comment{
Nothing good happens upon fixing the weak force and increasing the Mach
number $\ma$ of the initial flow (still keeping it subsonic). Now
the behavior of $\p$ gets similar to the $\f=0$ case. The work
extracted from the kinetic energy increases (due to a larger
$\kappa_1$), but it is now compensated by the negative work from the
enthalpy so that the overall work does not change much. In
contrast, the efficiency of work-extraction decreases due to a larger
$\ma$. }

\begin{figure}[ht]
\includegraphics[width=7cm]{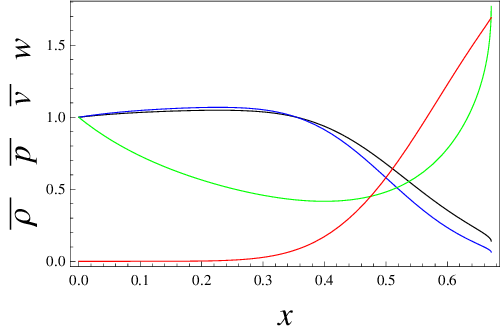}
\caption{ 
Dimensionless density $\r$ (black curve), pressure $\p$ (blue curve), 
velocity $\v$ (green curve), and work $w=\w/10$ (red curve)
versus $x$ obtained from solving (\ref{mora}, \ref{gugu})
for $x\in [0,0.6714]$ under the same parameters as in Fig.~\ref{fig1} but stronger force $f=1$.
The dimensionless velocity $\v(x)$ increases for $x>0.45$ reaching the sonic
value (\ref{maxim}) at the end point $x=x_2=0.6714$, where the work attains 
its maximum (\ref{maximo}). Cyclic value (\ref{mito}), where bound (\ref{balto}) is attained: $\r(x_2'=0.3547)=1$. 
}
\label{fig2}
\end{figure}

\comment{For situations exemplified in Fig.~\ref{fig2}, we cannot
continue the set-up for $x>x_2$ (by keeping the shapes of $\a(x)$ and
$\f(x)$ intact), because the sonic velocity $\v(x_2)=\v_{\rm 2\, m}$
[cf.~(\ref{durban})] is a singularity of (\ref{drob}). Appendix
\ref{sona} explains this aspect in detail.}

{\it Maximal work-extraction.} Eqs.~(\ref{maxim}, \ref{maximo}) show
that in the quasi-1d case ($\sigma=\v_{\rm tr}=0$) the maximal
work-extraction $\w_{\rm m}>0$ demands a positive contribution $\propto 1-\ma
\v^2_{\rm m}$ from enthalpy due to $\ma<1$ (subsonic input) and $\a_2>1$
(expanding area). Fig.~\ref{fig2} shows that $\w_{\rm m}$ is attained
in a strongly-forced quasi-1d case with a nonmonotonic $\v(x)$
[cf.~(\ref{shavkat})] that reaches the sonic value $\v(x_2)=\v_{\rm
m}>1$ in (\ref{maxim}). Hence the kinetic energy increases, a typical
scenario of attaining $\w_{\rm m}$ under subsonic input; see \S 6 of
\cite{sup}. Since $\w_{\rm m}>0$ is extracted from enthalpy only, the
efficiency is redefined by normalizing the work to enthalpy input
[cf.~(\ref{eq:5}, \ref{eq:2}, \ref{vedma2})]:
\BEA
\label{guppi}
\eta=\frac{-\int{\rm d}V\, \vec{v}\cdotp\vec{F}  }{ a_1(\rho_1\varepsilon_1+p_1)v_1  }
=\frac{\w_{\rm m} (\g-1)\ma}{2}.
\EEA
Using $\v(x_2)=\v_{\rm m}\geq 1$, $\sigma>0$ and $\gamma>1$ we bound from (\ref{maximo}, \ref{guppi}) 
the efficiency $\eta$ at the maximal work-extraction from enthalpy [cf. \S 6 of \cite{sup} and (\ref{balto})]:
\BEA
\label{ragoro}
\eta\leq 1-\left[\, {\ma}\,{\a_2^{-2}}(x_2) \,\right]^{\frac{\g-1}{\g+1}}.
\EEA
This bound is smaller than one, because we consider the initially subsonic 
regime $\ma<1$, and because $\a_2>1$.

For Fig.~\ref{fig2} the efficiency at the maximal work is
$\eta(x_2=0.6714)=0.4833$.  The work-extraction
(\ref{feofan}--\ref{mito}) from kinetic energy can be defined here at a
smaller value $x_2=x_2'=0.3547$, where cyclic condition (\ref{mito})
holds, and hence the bound (\ref{balto}) is reached and reads:
$\w(x_2')=0.8185$. This is larger than $0.4833$, but the work extracted
at efficiency $0.8185$ [cf.~(\ref{kurdian})] is smaller than the maximal
value (\ref{maximo}); see Fig.~\ref{fig2}. This conflict between
maximizing the work {\it vs.} efficiency resembles that for
heat-engines, where Carnot's efficiency is larger than the efficiency at
the maximal work, which for certain models has Curzon-Ahlborn's shape
\cite{callen}; see \S 6 of \cite{sup}.

\comment{
since now only the initial flux of enthalpy is the resource of
work-extraction. Now the efficiency $0.4833$ of the enthalpic
scenario in Fig.~\ref{fig2} is lower than the
efficiency $0.971$ of the kinetic work-extraction regime in
Fig.~\ref{fig1}. 
The latter is larger than Betz's limit $0.59$.}

{\it Outlook.} Our results show that the problem of work-extraction in
fluid dynamics is far from being closed and has analogies
with heat-engine physics. Possible future directions for this research are
quantum windmills \cite{nano_motor_1,nano_motor_2} and work-extraction
from quantum flows.

\comment{
We found that it is typically the case that the efficiency (\ref{balto})
of work-extraction from kinetic energy is larger than the efficiency at the maximal
work (\ref{maxim}, \ref{maximo}). However, analytical comparison between 
(\ref{fi}, \ref{maxim}, \ref{maximo}) and (\ref{balto}) does not confirm this fact. 
}

\acknowledgements

I thank S. Fauve for discussions and for hospitality in 
\'Ecole Normale Sup\'eriore (Paris), where a part of this work was done. 
I was supported by SCS of Armenia, grants No. 18RF-015 and No. 18T-1C090
and by Foundation for Armenian Science and Technology (FAST).

\clearpage

\section{Supplementary Material}

Supplementary Material consists of 9 chapters (referred to via \S).
{\bf References to equations and figures of the main text are marked
bold}. \S 1 discusses momentum conservation.  \S 2 studies Betz's model
in great detail and explains why specifically it is inapplicable.  \S 3
discusses limitations of the model. \S 4 shows that the bound {\bf (19)}
applies to non-ideal gases. \S 5 and \S 6 discuss the work versus
efficiency in two scenarios of work-extraction for (respectively) weak
and strong force. \S 7 explains details of the sonic limit. \S 8
considers implications of the Bernoulli equation. \S 9 studies
work-extraction in a cylindric tube and makes relation with physics of
d'Alembert's paradox. 

\subsection{1. Conservation laws of momentum}
\label{momo}

\begin{figure}[ht]
\includegraphics[width=7cm]{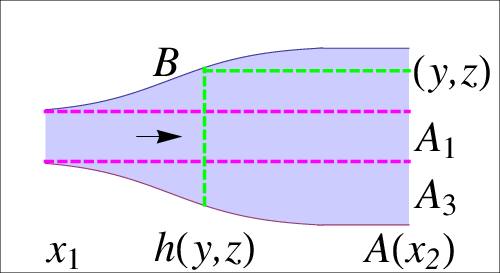}
\caption{Schematic representation of the integration domain in (\ref{veko}). 
$B$ is the control volume and $A(x)$ is its cross-section.
Magenta lines divide the output surface $A(x_2)=A_2$ over $A_1$ and $A_3$: $A_2=A_1\cup A_3$. 
$h(y,z)$ relates to a point $(y,z)\in A_3$; see (\ref{eq:17}). 
}
\label{fig22}
\end{figure} 

The conservation of momentum reads \cite{landau-1}
\BEA
  \label{eq:3}
  \vnab\cdot (\rho v_\ell\vec{v})+\partial_\ell p= {F}_\ell, 
\quad \ell=x,y,z,
\EEA
where $v_\ell$ and $F_\ell$ are components of (resp.) $\vec{v}$ and
$\vec{F}$.  Eq.~(\ref{eq:3}) is a combination of the mass conservation
and Euler's equation.  Eq.~(\ref{eq:3}) is considered separately from
other conservation laws, since within the general approach it does not
lead to useful constraints. Let us see why. 

Integrating (\ref{eq:3}) with $\ell=x$ over the volume $B$, and
using the same arguments as before {\bf (10)}, we get
\BEA
\rho_2a_2v_2^2-\rho_1a_1v_1^2+\int{\rm d}V\,\partial_xp(x,y,z)= 
\int{\rm d}V\,F_x.
  \label{eq:33}
\EEA
We need to treat
the volume integral $\int {\rm d}V\,\partial_xp(x,y,z)$ from
(\ref{eq:33}); see Fig.~\ref{fig22}. Given $(y,z)\in A(x_2)$,
let $h(y,z)$ be the minimal possible value of $x$; see Fig.~\ref{fig22}. Note that
$h(y,z)=x_1$, if $(y,z)\in A_1\subset A(x_2)$. Let also
$A_3$ be that part of $A_2=A(x_{2})$, which does
not project to $A_1$; see Fig.~\ref{fig22}. The integral is taken as follows
\begin{align}
  &\int{\rm d}V\,\partial_xp(x,y,z)= \int_{A_2}
  {\rm d}y\,{\rm d}z\int_{h(y,z)}^{x_2}{\rm d} x\,\partial_xp(x,y,z)\nonumber\\
  &= \int_{A_2} {\rm d}y\,{\rm d}z\, p(x_2,y,z) - \int_{A_{1}} {\rm d}y\,{\rm d}z\, p(x_{1},y,z)
  \nonumber\\
  & -\int_{A_{3}} {\rm d}y\,{\rm d}z\, p(h(y,z), y,z)   \label{veko}\\
  &= a_2p_2-a_{1}p_{1} -\zeta
  \label{eq:17}
\end{align}
where in (\ref{eq:17}) we employed {\bf (1-4)} and denoted 
\begin{gather}
  \label{eq:45}
\zeta \equiv
\int_{A_{3}} {\rm d}y\,{\rm d}z\, p(h (y,z), y,z)>0.
\end{gather}
Here $\zeta>0$ is due to $p>0$.

We now deduce from (\ref{eq:33}, \ref{eq:45}) and from the mass conservation $\rho_1v_1a_1=\rho_2v_2a_2$:
\BEA
\label{eq:03a}
v_{1} +\frac{p_{1}}{\rho_{1}v_{1}}+\frac{\zeta + \int{\rm d}V\, {F}_x}{a_{1}\rho_{1} v_{1}}
=v_{2} +\frac{p_{2}}{\rho_{2}v_{2}}.
\EEA
Eq.~(\ref{eq:03a}) is not useful, because it contains an unknown factor $\zeta$. 
More specifically, 
note that even the sign of the effective parameter $\zeta + \int{\rm d}V\, {F}_x$
is not fixed, since $\zeta>0$, while $\int{\rm d}V\, {F}_x<0$, as needed for work-extraction.

Likewise, we can consider (\ref{eq:3}) with $\ell=y$: 
\begin{gather}
a_2\rho_2v_2\,\left\langle\,\widetilde{\rho}(y,z)\,v_y(x_2,y,z)\,\right \rangle \nonumber \\
+\int {\rm d}V\,\partial_yp(x,y,z)=\int \d V\,{F}_y,
\label{ushi}
\end{gather}
where $\widetilde{\rho}(y,z)$ and $\left\langle ... \right \rangle$ are defined in (resp.) {\bf (3)} and {\bf (4)}.
Here as well, there is an unknown factor
$\int {\rm d}V\,\partial_yp(x,y,z)$. 

\subsection{2. Betz's model}
\label{hellobetz}

\begin{figure}[ht]
\includegraphics[width=7cm]{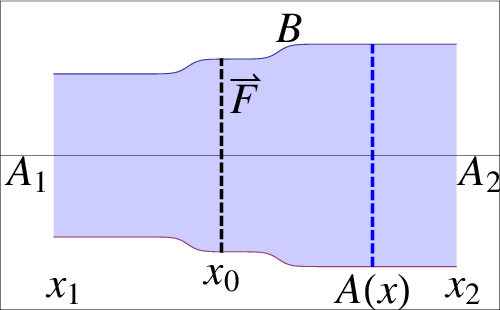}
\caption{Schematic representation of Betz's model. The flow (denoted by
blue) goes from $x_{1}$ (input) to $x_{2}$ (output). 
$\vec{F}=(F_x,0,0)$ is the external force that is localized in the vicinity of $x_0$. 
$B$ is the control volume (filled).  
$A(x)$ (dashed line) is the cross-section of $B$. 
$A_1=A(x_1)$ and $A_2=A(x_2)$ are (resp.) input and output surfaces. The area of $A(x)$ is $a(x)$.\\
The first difference with respect to the set-up shown in {\bf Fig.~1}
is that the force acts only within the cross-section $A(x_0)$ located at $x_0$; see (\ref{barba}). 
The second difference is that condition (\ref{bu3}) has to hold on $A(x_0)$, which 
constrains the surface of $B$ in the vicinity of $x_0$.
}
\label{fig11}
\end{figure} 

Here we study in detail the Betz's model that was reviewed in literature
several times
\cite{handbook-1,pelka-1,inglis-1,sorensen-1,greet-1,rauh-1}.  Similar
approaches were developed by Lanchester and Joukowsky; see
\cite{okulov-1} for details. 

Our conclusion will be that the model makes irrelevant assumptions and
that anyhow its conclusions do not concern the efficiency of
work-extraction from kinetic energy.  Not all of our negative
conclusions are new. The objection {\bf P1} (below in \S 2.3) was in
fact formulated in \cite{greet-1} though in a less explicit way.
Objection {\bf P2} was legitimately raised in \cite{rauh-1}. 

\subsubsection{2.1 Assumptions of the model}

We now spell out assumptions of the
model in great detail and put them in the context of conservation laws.

{\bf B0.} The flow model is shown in Fig.~\ref{fig11}.

{\bf B1.} The flow is dissipationless and incompressible
\BEA
\label{rhorho}
\rho_1=\rho_2=\rho(x)={\rm const}.
\EEA
Hence we can employ conservation of mass, energy and momentum only; cf.~{\bf (10, 11)} and (\ref{eq:03a}). 
No conservation of entropy is to be invoked, since the speed of sound is now infinite \cite{landau-1}. 

{\bf B2.} The external force $\vec{F}$ has zero transverse components, $F_y=F_z=0$, and
is localized in a thin domain around $x_0\in [x_1, x_2]$ [cf.~Fig.~\ref{fig11}]:
\begin{eqnarray}
  \label{barba}
  \vec{F}=(-f\,\,\delta(x-x_0),\,0,\,0)
\end{eqnarray}
where $\delta(x)$ is the delta-function, and where $f>0$ is a positive constant.

{\bf B3.}
The output pressure is homogeneous over input surface $A_1=A(x_1)$ 
and output surface $A_2=A(x_2)$ [cf.~{\bf (3)}]:
\BEA
p(x_1,y,z)=p_1,\\ p(x_2,y,z)=p_2.
\label{voch}
\EEA
One implication of (\ref{rhorho}, \ref{voch}) is that output transverse velocities are neglegible
due to Bernoulli's equation [see \S 7 below for details]:
\BEA
  \label{arba}
v_y(x_2,y,z)=v_z(x_2,y,z)=0.
\EEA
Alternatively, (\ref{arba}) can be taken as an additional assumption, as usually done in literature
\cite{handbook-1,pelka-1,inglis-1,sorensen-1,greet-1,rauh-1}. 

{\bf B4.} Input and output pressures are equal [cf.~(\ref{voch})]:
\BEA
\label{pp}
p_1=p_2=p.
\EEA
Due to (\ref{rhorho}, \ref{voch}, \ref{pp}) the work is extracted from the kinetic energy only, as confirmed below. 

{\bf B5.}
Assumptions expressed by Eq.~{\bf (1,2)} on the homogeneous axial 
velocity $v_x$ at input and output are naturally done also for Betz's 
model. 
Moreover, it is assumed that the analogue of {\bf (1,2)} hold as well at $A(x_0)$, 
i.e. altogether we have [cf.~Fig.~(\ref{fig11}]
\begin{gather}
\label{bu2}
v_x(x_1,y,z)=v_1, \qquad v_x(x_2,y,z)=v_2, \\ v_x(x_0,y,z)=v(x_0).
\label{bu3}
\end{gather}
Eqs.~(\ref{rhorho}, \ref{bu2}, \ref{bu3}) allow us to write the mass conservation as [cf.~{\bf (10)}]: 
\BEA
\label{orbit}
a_1v_1=a(x_0)v(x_0)=a_2v_2. 
\EEA

{\bf B6.} We employ (\ref{rhorho}, \ref{voch}, \ref{pp}) in 
momentum conservation relation (\ref{eq:03a}), and assume {\it additionally}
that the following relation takes place in (\ref{eq:03a}):
\BEA
\label{dod}
\frac{p}{v_{1}}+\frac{\zeta}{a_{1} v_{1}}=\frac{p}{ v_{2}},
\EEA
which via the mass conservation $a_1v_1=a_2v_2$ amounts to:
\BEA
\label{od}
\zeta=p(a_2-a_1).
\EEA
Assumption (\ref{dod}) allows to fix the unknown $\zeta$.

Looking at definition (\ref{eq:45}) of $\zeta$, we can replace
(\ref{od}) by an assumption that pressure is constant over the whole
surface of the volume $B$ between $A(x_1)$ and $A(x_2)$; cf.~(\ref{pp}).
Assumption (\ref{od}) is normally made implicitly
\cite{handbook-1,pelka-1,inglis-1,sorensen-1,greet-1}. It was spelled
out explicitly in \cite{rauh-1}.

\subsubsection{2.2 Derivation of Betz's limit}

\comment{Eq.~(\ref{eq:17}) shows that (\ref{pp})
and (\ref{od}) amount to assuming that $\int_{V_{1|2}}{\rm d}V\,\partial_xp(x,y,z)=0$. }

Recall the energy conservation law {\bf (8)}, where in view of incompressibility assumption (\ref{rhorho}) we
should skip the internal energy $\varepsilon$ \cite{landau-1}:
\BEA
\label{eq:222}
\vnab\cdot\left[\frac{\rho \vec{v}^2\vec{v}}{2}+p\vec{v}\right]  =\vec{v}\cdot\vec{F}.
\EEA
Then instead of {\bf (11)}  we get from (\ref{pp}, \ref{orbit}, \ref{eq:222}):
\BEA
-\frac{\int {\rm d}V\, \vec{v}\cdot\vec{F}  }{a_{1}\rho v_{1}}=  \frac{v^2_{1}-v^2_{2}}{2},
\label{ursula}
\EEA
where the transverse velocity contribution in (\ref{ursula}) is already skipped 
due to (\ref{arba}), and we already employed (\ref{pp}). Eq.~(\ref{ursula}) 
makes it clear that the work is extracted from kinetic energy only. 
Using (\ref{barba}, \ref{orbit}) in (\ref{ursula}) we get
\BEA
\label{dublin}
\frac{v_1^2}{2}-\frac{v_2^2}{2}=\frac{a(x_0)v(x_0)f}{a_1v_1\rho}=\frac{f}{\rho}.
\EEA
Likewise, (\ref{eq:03a}) leads together with (\ref{od}), (\ref{barba}) and (\ref{rhorho}, \ref{pp}):
\BEA
\label{berlin}
v_1-v_2=\frac{a(x_0)f}{a_1v_1\rho}.
\EEA
Eqs.~(\ref{orbit}, \ref{dublin}, \ref{berlin}) imply Drude's relation for (\ref{bu3}) \cite{drud}:
\BEA
\label{drude}
v(x_0)=(v_1+v_2)/2.
\EEA

Using (\ref{barba}, \ref{dublin}, \ref{drude}) we write for the work:
\BEA
&&-\int {\rm d}V\, \vec{v}\cdotp\vec{F}=fv(x_0) a(x_0)\nonumber\\
&&=\rho a(x_0) \frac{(v_1+v_2)(v_1^2-v_2^2)}{4}\nonumber\\
&&=\frac{v_1^3}{4}\rho a(x_0)(1+\v_2)(1-\v_2^2),
\label{dam}
\EEA
where $\v_2\equiv v_2/v_1$. 
Now consider the ratio
\BEA
\frac{-\int {\rm d}V\, \vec{v}\cdotp\vec{F}}{\frac{1}{2}\rho a(x_0)v_1^3 }
=\frac{1}{2}(1+\v_2)(1-\v_2^2).
\label{damdam}
\EEA
The RHS of (\ref{damdam}) maximizes for 
\BEA
\label{zoro}
\v_2=v_2/v_1={1}/{3}, 
\EEA
leading from (\ref{dam}) to the Betz's (upper) limit for the ratio in the LHS of (\ref{damdam}):
\begin{gather}
\label{bebe}
\frac{-\int {\rm d}V\, \vec{v}\cdotp\vec{F}}{\frac{1}{2}\rho a(x_0)v_1^3 }
\leq \frac{16}{27}.
\end{gather}

\comment{
Below we show that the assumptions leading to (\ref{dam}) are not
consistent with conservation laws, because making only some of those
assumptions leads to an expression explicitly different from
(\ref{dam}). As a consequence of this fact, the bound (\ref{bebe}) is
irrelevant, i.e. it may or may not satisfied for occasional reasons. 
}

\comment{
We can derive a more general result from weaker assumptions.
If instead of (\ref{rhorho}) we assume only $\rho_1=\rho_2$ and also
neglect the transverse contribution $\hat u=0$, then together with (\ref{zoro}) a slightly
generalized versions of (\ref{dam}, \ref{bebe}) are derived:
\BEA
\label{dam1}
-\int {\rm d}V\, \vec{v}\cdotp\vec{F}&=&\frac{v_1^3}{4}\rho(x_0)a(x_0)(1+\v_2)(1-\v_2^3)~~~\\
\label{bebe1}
&\leq& \frac{\rho(x_0)a(x_0) v_1^3}{2} \times \frac{16}{27}.
\EEA

Note that Ref.~\cite{greet} criticized assumptions of Betz's model, and
made instead its own assumptions. In particular, it was assumed that the
velocity is a smooth function around $x_0$. Further assumptions of
\cite{greet} are unclear for us, which agrees with the opinion of
Ref.~\cite{rauh}. 
}

\subsubsection{2.3 Problems of Betz's model}

{\bf P1.} In (\ref{bebe}), $\frac{16}{27}$ is now interpreted as an
upper limit of the efficiency for the work-extraction from kinetic
energy \cite{handbook-1,pelka-1,inglis-1,sorensen-1,greet-1}.  
This is not correct. The correct efficiency $\eta$ of the work
extraction from kinetic energy is defined as the work divided over the
influx of kinetic energy:
\BEA
\label{dam1}
\eta =
\frac{-\int {\rm d}V\, \vec{v}\cdotp\vec{F} }{\frac{1}{2}a_1\rho v_1^3}.
\EEA
We remind that the LHS of (\ref{bebe}) cannot be an efficiency, also
because it came out of idealization (\ref{barba}). If the force were not
artificially localized around $x_0$ [as (\ref{barba}) does], the surface
$A(x_0)$ would not have any specific meaning \cite{choban}.  Once the
correct quantity (\ref{dam1}) is employed the above derivation becomes
pointless. Indeed, we return to to (\ref{dam}) and note that upon using
(\ref{orbit}, \ref{drude}), $a(x_0)=\frac{2a_1}{1+\v_2}$ can also be
presented as a function of $\v_2$. Hence for the correct efficiency
$\eta$ we get from (\ref{dam}): $\eta=1-\v_2^2$, whose upper limit is
just $1$. 

One can try to apply (\ref{bebe}) to $\eta$ assuming $a(x_1)\simeq a(x_0)$. 
This assumption is untenable, because then the mass conservation law
(\ref{orbit}) implies $v(x_1)\simeq v(x_0)$, and then (\ref{drude})
leads to $v_1\simeq v_2$.

{\bf P2.} Motivations for assuming (\ref{od}) are unclear.  A rationale
for (\ref{od}) can relate to boundary conditions $\vec{v}=0$ on the
surface ${\cal B}$ of the volume $B$ (see Fig.~\ref{fig11}), which from the
Euler equation leads to $\vec{\nabla} p|_{{\cal B}}=0$ and hence to
$p(\vec{r})={\rm const}$ for $\vec{r}\in {\cal B}$. However, $B$ is defined
via flow lines, i.e. the boundary condition $\vec{v}|_{{\cal B}}=0$ 
does not apply. Moreover, when Betz's model is presented graphically one notes
that the pressure is non-monotonic and depends only on $x$
\cite{pelka-1,inglis-1,sorensen-1,greet-1,rauh-1}. Hence making
it constant for the whole surface between $A(x_1)$ and $A(x_2)$ is an
arbitrary assumption. 

{\bf P3.} In the main text we implemented some assumptions of the Betz's
model and got different results.  In particular, assumptions
(\ref{voch}, \ref{arba}) of the Betz's model lead to $\sigma=v_{\rm
tr}=0$. Using these in {\bf (17)}|which is necessary and sufficient for
the work-extraction from kinetic energy|we get
\BEA
\label{ugo}
v_2=v_1a_1/a_2.
\EEA
The work is then extracted from the kinetic energy only and amounts to
{\bf (18, 19)}
\BEA
\label{chaves}
-{\int {\rm d}V\, \vec{v}\cdotp\vec{F}  }=\frac{\rho v_1^3 a_1}{2}(1-\frac{a_1^2}{a^2_2}).
\EEA
It is seen that (\ref{ugo}) and (\ref{chaves}) are different from
(respectively) (\ref{zoro}) and (\ref{bebe}). In particular,
(\ref{chaves})|in contrast to (\ref{bebe})| refers to the efficiency of
work-extraction. Hence assumptions of
the Betz's model are not consistent with each other, since we
implemented some of them into the conservation laws and got different
results. 

\subsubsection{2.4 Limits on Betz's efficiency}

Let us call the LHS of (\ref{bebe}) Betz's efficiency:
\BEA
\zeta_{\rm B}\equiv \frac{-\int {\rm d}V\, \vec{v}\cdotp\vec{F}}{\frac{1}{2}\rho a(x_0)v_1^3 }.
\EEA
Since $a(x_0)\geq a(x_1)$ ($a(x_1)=a_1$ is the input area), $\zeta_{\rm B}$ is smaller than the actual efficiency:
\BEA
\zeta_{\rm B}=\frac{a(x_1)}{a(x_0)}\eta\leq \eta.
\EEA
Hence when the efficiency $\eta$ holds the bound {\bf (19)}, $\zeta_{\rm B}$ will hold:
\BEA
\zeta_{\rm B}\leq\frac{a(x_1)}{a(x_0)}\left(1-\frac{a^2(x_1)}{a^2(x_2)}   \right).
\EEA

\subsubsection{2.5 Conclusion on Betz's model}

The model together with its assumptions is problematic for several
reasons. But it did motivate the development of the model in the
main text. 

\subsection{3. Limitations of the model and open problems}

\subsubsection{3.1 Remarks on the general structure of the model }

The most general approach for solving the model pictured in {\bf Fig.~1}
is to give the shape of the force $\vec{F}$, provide boundary conditions
at the input (which are homogeneous for the present model), and to
determine the 3d flow in the full space by solving the (compressible)
Euler equation together with continuity equations for mass and entropy.
In particular, such a solution will determine the input and output
cross-section areas $a_1=A(x_1)$ and $a_2=A(x_2)$ of the control volume
$B$; see {\bf Fig.~1}.  Recall that the choice of $B$ has to hold the
following conditions (see after {\bf 1} in the main text): {\it (i)} integration over $B$ 
suffices for calculating the total work done by $\vec{F}$:
\BEA
\int_\Omega \d^3
r\, (-\vec{v}\cdotp\vec{F})=\int\d^3 r\, (-\vec{v}\cdotp\vec{F})
=\int_B \d^3 r\, (-\vec{v}\cdotp\vec{F}).
\EEA
{\it (ii)} The input surface $A_1=A(x_1)$ is
possibly small. This is needed for ensuring that the output surface
$A(x_2)$ is small as well (hence $v_x$ can be homogeneous on it, see {\bf (2)}),
and also for maximizing the efficiency of work-extraction, where the
work is divided on the area $a_1$ of $A(x_1)$. 

The general approach is not practical, since the 3d solution is
certainly not available for any sufficiently non-trivial $\vec{F}$.
Instead, the main text introduced the control volume $B$, applied to $B$
conservation laws of mass, energy and entropy. These led to upper
bounds for the efficiency {\bf (19)} of work-extraction from kinetic
energy, and for the efficiency at the maximal work (extracted from
enthalpy); see {\bf (35)}. 

These expressions are universal in the sense that they do not depend on
details of $\vec{F}$. {\bf Eq.~(19)} also does not depend on the input
characteristics of the flow, and on the assumed ideal gas feature of the
fluid. {\bf Eq.~(35)} depends on the adiabatic index of the ideal gas
and on the initial Mach number of the flow. 

Before applying these bounds in practice, one needs to estimate the
input and output areas $a(x_1)$ and $a(x_2)$. The situation here is
similar to applying the Carnot bound $1-(T_{\rm cold}/T_{\rm hot})$ to
realistic heat-engines, e.g. to internal combustion engines. Here
$T_{\rm cold}$ is given as the atmosphere temperature, but $T_{\rm hot}$
depends on the very functioning of the engine, since this is the
temperature that is created by the combustion process. Hence estimating
$T_{\rm hot}$ demands a knowledge of the heat-engine functioning.

\comment{
{\it (1.)} In the main text we assumed that the force $\vec{F}$ and the
turbine shape $B$ are given (not necessarily independently); see {\bf Fig.~1}
(of the main text).  This is inspired by the quasi-1d approach, where the
flow is determined by the force and shape as two independent inputs
\cite{shirokov-1,landau-1}.  Then we looked for various physical
conditions to be imposed on the flow features. First it assumes that it
is possible to extract work from the kinetic energy only;
cf.~Eq.~{\bf (17)}. This assumption does have an empirical support, and
it holds within the quasi-1d approach. }

\subsubsection{3.2 Limitations of concrete assumptions on the flow}

-- Assumption {\bf 2} on a stationary force $\vec{F}(\vec{x})$ 
is restrictive, because wind turbines have blades that move faster than
wind \cite{sorensen-1}, and do not just exert a stationary force on the flow.

-- Assumptions {\bf 3} and {\bf 5} in the main text are limited by
turbulence \cite{turbu_1,turbu_2}, since the turbulence makes the
velocity time-dependent and space-dependent (i.e. inhomogeneous).
However, after taking time-averages, the homogenuity is frequently
recovered \cite{shirokov-1,landau-1,pelka-1}, and together with it
assumptions {\bf 4} and {\bf 5} are supported. 

-- Assumption {\bf 3} is also limited by vorticity of the input flow.
Excluding input vorticity seems legitimate if we want
to understand work-extraction from the simplest form of kinetic
energy. Clearly, vorticity is a separate resource for work-extraction
and should be studied in future for its own sake.

-- Assumption {\bf 5} is standard for quasi-1d and/or hydraulic flows;
see \cite{luchini-1,manneville-1,noble-1} for recent expositions and reviews,
earlier literature on the subject is reviewed in
\cite{hicks-1,yen-1,shirokov-1,sedov-1}. (Note that assumption {\bf 5} and
{\bf (2)} is not limited to the quasi-1d situation, since it only
concerns the longitudinal component of the flow; i.e. tangential
components need not nullify; cf.~{\bf (2)}.) The assumption is feasible for ideal
(dissipationless) fluid, since it is consistent with the corresponding
boundary conditions, i.e. the normal components to surface nullifies;
see {\bf Fig.~1}. It is less suitable for viscous fluid in relatively
thin pipes and ducts (as well as in open flow), but even for such cases
the deviations from it are well-controllable and frequently small, as
experiments show \cite{franz}.  Thus ssumption {\bf 5} does have both
empirical and theoretical support. Moreover, it is known what one can do
when it does not hold, e.g. introduce additional variables or improve
the velocity behavior next to boundaries.  Unfortunately, all
(improving) works rely on the incompressibility assumption
\cite{luchini-1,manneville-1,noble-1,hicks-1,yen-1} and hence do not apply
directly to the considered situation, where the compressibility can be
small, but instrumental. At least some of them ought to be generalizable
to the compressible case, i.e. it should be possible to improve on
assumption {\bf 5} and explore situations, where it does not hold. 

-- Assumption {\bf 6} on the ideal gas feature of the fluid is partially 
relaxed in \S 4 below.

-- Note that {\bf (16)} assumes that for the purpose of maximizing the
(dimensionless) work, the final dimensionless velocity $\bar{v}_2$ can
be varied independently from other parameters involved in {\bf (16)}:
$\bar{v}_{\rm tr}^2$, $\sigma$, $\a_2$ and $\ma$. This assumption does
make sense for the following reasons. First, if there are relations
between these parameters, then the maximal work will be smaller. I.e.
the expression obtained in {\bf (21, 22)} via the unconditional
maximization still provides an upper bound on the work. Eqs.~{\bf (21,
22)} allow to conclude on the sonic character of optimal output
velocities. Second, this assumption is confirmed in the quasi-1d
approach.

\subsection{4. Efficiency of work-extraction from kinetic energy
holds {\bf (18)} for non-ideal gases}
\label{fernan}

Eqs.~{\bf (18, 19)} show that the efficiency of work extraction from
kinetic energy of an ideal-gas flow holds an upper bound {\bf (19)}. The
ideal-gas is understood in terms of {\bf (5)}.  Eq.~{\bf (18)} shows
that the real efficiency is always smaller due to inhomogeneous output
pressure and density; see {\bf (4)}. 

Here we relax the assumption {\bf (5)} on the ideal gas, but are able to
prove a more restrictive statement: if we assume that the output
pressure and density are homogeneous, then the efficiency of work
extraction from kinetic energy is given by {\bf (19)}.  Put differently,
we were not able to show that for non-ideal gases inhomogeneous output
pressure or density decreases the efficiency. 

Now conservation of entropy and mass amount to
\BEA
s(p_1,\rho_1)=s(p_2,\rho_2),
\label{boka}
\EEA
while the fact that no work is extracted from enthalpy reads
\BEA
\label{vuk}
\psi(p_1,s(p_1,\rho_1))=\psi(p_2,s(p_2,\rho_2)),
\EEA
where $\psi=\varepsilon+\frac{p}{\rho}$ is the enthalpy density, and where in (\ref{vuk}) 
we recalled that natural variables of $\psi$ are $p$ and $s$. Now (\ref{boka}, \ref{vuk})
imply $p_1$=$p_2$ from $\psi(p_1,s)=\psi(p_2,s)$, because $\partial\psi/\partial p|_s=1/\rho>0$. 
It remains to show that $s(p,\rho_1)=s(p,\rho_2)$ [cf.~(\ref{boka})] has the only solution $\rho_1=\rho_2$.
This will be shown via demonstrating that $\partial s/\partial \rho|_p$ has a constant sign. Employing
thermodynamic inequalities we show below that 
\BEA
\label{glob}
{\rm sign}[\,\partial s/\partial \rho|_p\,]=-{\rm sign}[\,\partial p/\partial T|_\rho\,].
\EEA
Now for many cases of practical interest one can demonstrate
\BEA
\label{ar}
{\rm sign}[\,\partial p/\partial T|_\rho\,]>0, 
\EEA
directly from the equation of state. Though (\ref{ar}) is not among standard thermodynamic inequalities, 
we are not aware of any realistic example, where it is violated. 
Here is the example of the van der Waals gas, where it holds:
\begin{gather}
p=\frac{\vartheta\rho}{1-\rho b}-\rho^2 a,\qquad \vartheta\equiv RT/\mu,
\end{gather}
where $a>0$ and $b>0$ are the van der Waals parameters, 
$T$ is temperature, $R$ is the gas constant and $\mu$ is the molar mass \cite{landau-1}. 
Recall that $1>\rho b$ is a strict constraint for the van der Waals gas \cite{landau_stat-1}.

Once (\ref{boka}, \ref{vuk}) are solved only for $p_1=p_2$ and
$\rho_1=\rho_2$ (cyclicality conditions), we employ the conservation of
mass $a_1\rho_1v_1=a_2\rho_2v_2$ to show that the efficiency of
work-extraction from kinetic energy equals {\bf (19)}.

To show (\ref{glob}), we focus on 
$\partial s/\partial V|_p$ (derivative of entropy over 
volume at fixed pressure), and write in natural thermodynamic variables $(V,T)$:
\BEA
\label{kur1}
\d s= \left.\frac{\partial s}{\partial V}\right|_T\d V+\left.\frac{\partial s}{\partial T}\right|_V\d T.
\EEA
Likewise the equation of state $p=p(V,T)$ implies
\BEA
\label{kur2}
\d p= \left.\frac{\partial p}{\partial V}\right|_T\d V+\left.\frac{\partial p}{\partial T}\right|_V\d T.
\EEA
Now a constant pressure implies $\d p=0$ in (\ref{kur2}). Employing this in (\ref{kur1}) we get
\BEA
\label{kur3}
\left.\frac{\partial s}{\partial V}\right|_p=\left.\frac{\partial s}{\partial V}\right|_T-\left.\frac{\partial s}{\partial T}\right|_V\times
\frac{\left.\frac{\partial p}{\partial V}\right|_T  }{\left.\frac{\partial p}{\partial T}\right|_V }.
\EEA
Employing Maxwell's relation $\left.\frac{\partial p}{\partial T}\right|_V=\left.\frac{\partial s}{\partial V}\right|_T$ \cite{landau_stat-1}, the fact
of $\left.\frac{\partial s}{\partial T}\right|_V=\frac{c_V}{T}>0$ (the constant volume heat-capacity is positive due to a thermodynamic inequality),
and $\left.\frac{\partial p}{\partial V}\right|_T<0$ (another known thermodynamic inequality) \cite{landau_stat-1} we conclude from (\ref{kur3})
that $\left.\frac{\partial s}{\partial V}\right|_p$
and $\left.\frac{\partial p}{\partial T}\right|_V$ have the same sign. Eq.~(\ref{glob}) follows from here upon noting $V=1/\rho$ and hence
\BEA
\left.\frac{\partial s}{\partial V}\right|_p=-\rho^2\, \left.\frac{\partial s}{\partial \rho}\right|_p.
\EEA

\begin{figure}[ht]
\includegraphics[width=8cm]{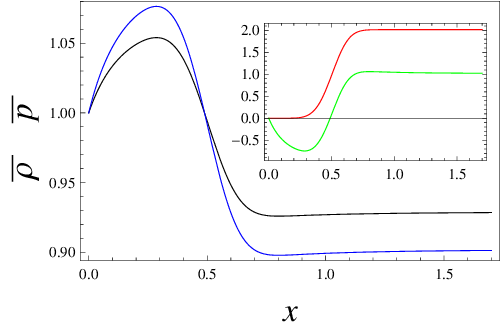}
\caption{This figure demonstrates the conflict between maximization of the extracted work and maximization of efficiency. \\
The main figure shows dimensionless density $\r$ (black curve) and pressure $\p$ (blue curve)
versus $x$. The inset shows the dimensionless work $\w$ (red curve), see {\bf (16)}, and the enthalpic part $\w_{\rm ent}$ 
given by (\ref{py}). All these quantities are obtained from solving {\bf (25, 26)}. \\
Parameters are those of {\bf Fig.~2} (of the main text).
But now $f=0.2$, i.e. the force is two times stronger than in {\bf Fig.~2}.
}
\label{fig1.1}
\end{figure} 

\subsection{5. Weak force: conflict between maximization 
of the extracted work and maximization of efficiency}

Eq.~{\bf (19)} deduces an upper bound for the efficiency of
work-extraction assuming that the work is extracted from kinetic energy
and the enthalpic contribution in {\bf (16)}:
\BEA
\label{py}
\w_{\rm ent}=\frac{2}{\ma(\g-1)}(1-e^{\sigma}\a_2^{1-\g}\,\v_2^{1-\g}),
\EEA
is precisely zero: $\w_{\rm ent}=0$. Eq.~{\bf (19)} does not refer to
maximizing the work and it is interesting to see how the efficiency
changes when the work is optimized over the choice of the end-point
$x_2$ for a given force $F$ in {\bf (25, 26)}. 

Fig.~\ref{fig1.1} studies the same situation as {\bf Fig.~2}, but now
$f=0.2$, i.e. the force is two times stronger. It is seen from
Fig.~\ref{fig1.1} that if we choose $x_2=0.49$, then the process is
cyclic, i.e. $\r_1=\r_2$ and hence $\p_1=\p_2$; cf.~{\bf (20)}.  Then
the enthalpy does not contribute to the work, and since for the
considered quasi-1d situation we have $\sigma=\v_{\rm tr}=0$, then the
efficiency of work-extraction from the kinetic energy is given by its
value {\bf (19)}. With parameters of Fig.~\ref{fig1.1} this value
amounts to $0.8897$. 

But with this choice of $x_2$ we loose nearly the half of the available
work, as shown by the read curve (for $\w$) and the green curve (for
$\w_{\rm ent}$) in Fig.~\ref{fig1.1}.  Choosing a larger value of $x_2$,
i.e. $x_2>0.7$, we shall increase the extracted work, but now the work
comes both from the enthalpy and kinetic energy, i.e. $\w_{\rm ent}>0$
in (\ref{py}). This fact implies that the efficiency of work-extraction
should be redefined, i.e.  it is now given as the ratio of the work to
the full input of energy [cf.~{\bf (8, 33)}]:
\BEA 
\label{effa}
&&\frac{-\int {\rm d}V\, \vec{v}\cdotp\vec{F}  }
{a_1(\rho_1\varepsilon_1+p_1)v_1+\frac{1}{2}a_{1}  \rho_{1} v^3_{1}}
=\frac{-\int {\rm d}V\, \vec{v}\cdotp\vec{F}  }{\gg a_1 p_1v_1+\frac{1}{2}a_{1}  \rho_{1} v^3_{1}}\nonumber\\
&&=(\frac{2}{(\gamma-1)\ma}+1)^{-1}\w. 
\EEA
For parameters of Fig.~\ref{fig1.1} the value of this redefined
efficiency (\ref{effa}) is $0.0555$, which is expectedly (much)
smaller than the efficiency $0.8897$ obtained above.

We conclude that there is generally a conflict between maximizing the
efficiency and maximizing the work.  The core of this conflict is that
the work can be increased due to a contribution from enthalpy. This fact
leads to redefining|and thereby decreasing|the efficiency. Such a
conflict need not be present always, i.e. it is absent for
parameters employed in {\bf Fig.~2}. 

\begin{figure}[ht]
\includegraphics[width=7cm]{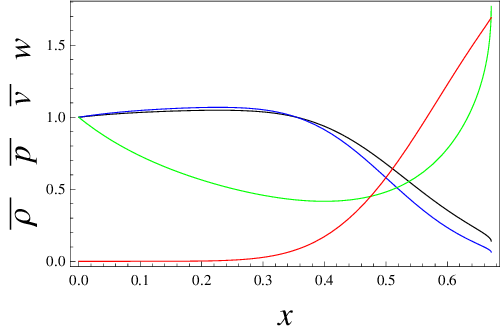}
\caption{ This figure demonstrates the reachability of the maximum work for a strong force.\\
Dimensionless density $\r$ (black curve), pressure $\p$ (blue curve), 
velocity $\v$ (green curve), and work $w=\w/10$ (red curve)
versus $x$ obtained from solving {\bf (25, 26)} (of the main text)
for $x\in [0,0.6714]$ under the same parameters as in {\bf Fig.~2} (and Fig.~\ref{fig1.1}), 
but now $f=1$, i.e. the force is stronger.\\
The dimensionless velocity $\v(x)$ increases for $x>0.45$ reaching the sonic
value {\bf (21)} at the end point $x=x_2=0.6714$, where the work attains its maximum {\bf (22)}. 
Cyclic values: $\r(x_2'=0.3547)=1$, $\v(x_2^*=0.6331)=1$.
}
\label{fig2.1}
\end{figure} 

\subsection{6. Strong force: the maximal work extraction}

\subsubsection{6.1 Upper bound on the efficiency of the maximal work-extraction}

Let us now turn to discussing the work-extraction in the strong-force
regime, which in the main text is exemplified by {\bf Fig.~3}. For
convenience this figure is reproduced as Fig.~\ref{fig2.1}. Let us repeat
the maximal work expressions {\bf (21, 22)} as
\begin{gather}
\label{maxim-1}
\v_2=\v_{\rm m}=\left( e^\sigma \a_2^{1-\g}{\rm M}_1^{-2} \right)^{\frac{1}{\g+1}},\\
\w_{\rm m}=\w(\v_{\rm m})=1-\v_{\rm m}^2-\v^2_{\rm tr}+\frac{2(1-\ma\,\v_{\rm m}^2)}{(\g-1)\ma},
\label{maxim-2}
\end{gather}
where the last equality can be also rewritten as
\begin{gather}
\w_{\rm m}=1-\v_{\rm m}^2-\v^2_{\rm tr} +\frac{2\left[ 1-e^{\frac{2\sigma}{1+\g}}
({\ma}\,{\a_2^{-2}} )^{\frac{\g-1}{\g+1}} \right]}{(\g-1)\ma}.
\label{maximo-2}
\end{gather}
If the maximal work-extraction $\w_{\rm m}>0$ takes place from enthalpy only (not from kinetic energy),
i.e. if in (\ref{maximo-2}) 
\BEA
\label{sudan}
1-\v_{\rm m}^2-\v^2_{\rm tr}\leq 0,
\EEA
then the efficiency of work-extraction 
is obtained by analogy to (\ref{effa}), where only the
influx of entahlpy is to be retained, since it is now the only resource
[cf.~{\bf (33)}]:
\BEA
\label{guppi-1}
\eta=\frac{-\int{\rm d}V\, \vec{v}\cdotp\vec{F}  }{ a_1(\rho_1\varepsilon_1+p_1)v_1  }
=\frac{-\int{\rm d}V\, \vec{v}\cdotp\vec{F}  }{ \gg a_1 p_1v_1  }
=\frac{\w\, (\g-1)\ma}{2}.
\EEA 
Using (\ref{sudan}), $\sigma>0$ and $\gamma>1$ we obtain from (\ref{maximo-2}, \ref{guppi-1}) 
the following upper bound for the efficiency $\eta$ at the maximal work-extraction from enthalpy:
\BEA
\label{ragor}
\eta\leq 1-\left[\, {\ma}\,{\a_2^{-2}}(x_2) \,\right]^{\frac{\g-1}{\g+1}}.
\EEA
Let us compare features of (\ref{ragor}) with bound
{\bf (19)} for the efficiency of work-extraction from kinetic energy.

--The meaning of (\ref{ragor}) is that a positive work is extracted from
enthalpy in the maximum work regime.  The contribution of the kinetic
energy to work is non-positive. The meaning of {\bf (19)} is that work
is extracted from the kinetic energy, when the enthalpic contribution is
zero. Bound {\bf (19)} cannot be derived if we demand that the enthalpic
contribution is non-positive, and the positivity of work is due to
kinetic energy. This is the reason why {\bf (19)}|in contrast to
(\ref{ragor})|relates to cyclic processes. 

--Eq.~(\ref{ragor}) is non-trivial (i.e. smaller than one) under $\ma\a_2^{-2}<1$.
This inequality is ensured by our consideration, since we consider
the initially subsonic regime $\ma<1$ ($\g>1$ for thermodynamic
reasons), and because $\a_2>1$. Eq.~{\bf (19)} is non-trivial under $\a_2>1$ only.

--Bound (\ref{ragor}) is similar to bound {\bf (19)} for the efficiency
of work-extraction from kinetic energy, because both (\ref{ragor}) and
{\bf (19)} do not depend on details of the force $\vec{F}$. But in
contrast to {\bf (19)}, bound (\ref{ragor}) depends also on the initial
Mach number $\ma$ and on the adiabatic index $\gamma$ of the fluid ideal
gas. The latter fact is natural, since (\ref{ragor}) refers to
work-extraction from enthalpy. 

--Yet another difference between (\ref{ragor}) and {\bf (19)} is that
(\ref{ragor})|in contrast to {\bf (19)}|is attainable under a rather
restrictive condition $1-\v_{\rm m}^2-\v^2_{\rm tr}=0$, which for the
quasi-1d situation transforms to $1-\v_{\rm m}^2=0$. This condition does
not generally hold. 

--When comparing (\ref{ragor}) with {\bf (19)} within the same set-up,
we should recall that $\a_2$ in {\bf (19)} and $\a_2(x_2)$ in
(\ref{ragor}) refer to different choices of $x_2$, as determined from
{\bf (17)} and (\ref{sudan}), respectively. 

Another general conclusion follows from (\ref{maximo-2}) upon noting
that in the quasi-1d situation we get $\v_{\rm tr}=\sigma=0$, i.e. in
the maximal work-extraction regime some (positive) work should come also
from enthalpy. 

\subsubsection{6.2 Numerical studies }

Fig.~\ref{fig2.1} shows that the maximal work (\ref{maxim-2}) is reached
for a subsonic, quasi-1d flow; hence we should put $\sigma=\v_{\rm
tr}=0$ in (\ref{maxim-1}--\ref{maximo-2}).  The dimensionless velocity
$\v(x)$ is a non-monotonic function of $x$: first it decays, as expected
due to expanding domain $\a(x)$, but then it starts to increase and
reaches for $x_2=0.6714$ the sonic value (\ref{maxim-1}) [cf.~{\bf
(23)}]; see Fig.~\ref{fig2.1}. Such a non-monotonic behavior is in
accord with {\bf (31, 32)} in the strong-force situation.  The work
grows and reaches the maximal value (\ref{maxim-2}) at the interval-end
$x_2=0.6714$; see Fig.~\ref{fig2.1}.  We remind that this is the maximal
possible value of work for a fixed input Mach number ${\rm M}_1$,
$\gamma$ and $\a(x_2)$.  The solution of {\bf (25, 26)} shown in
Fig.~\ref{fig2.1} cannot be continued for $x>x_2$, because the sonic
value of the velocity is a singularity point \cite{landau-1}; see \S 7
for details.

Note from Fig.~\ref{fig2.1} that there is another choice of $x_2$:
$x_2=x_2'=0.3547$, where the cyclic condition holds:
$\r(x_2'=0.3547)=1$, and hence $\p(x_2'=0.3547)=1$; cf.~{\bf (20)}.
Under this choice we return to the work extracted from kinetic energy.
The efficiency is given by {\bf (19)} (i.e. the bound in {\bf (19)} is
reached), which for parameters of Fig.~\ref{fig2.1} reads
\BEA
\label{pashto}
\w(x_2'=0.3547)=0.8185.
\EEA
But it is clear from Fig.~\ref{fig2.1} that at $x_2=x_2'=0.3547$ the 
work is far from its maximal value. Taking $x_2\in (0.3547, 0.6333)$
will lead to efficiencies sizably lower than (\ref{pashto}), because the
work is now extracted both from kinetic energy and enthalpy; cf.~(\ref{effa}). 
However for 
\BEA
\label{hazare}
\v_2(x)>\v_2(x_2^*=0.6333)=1 
\EEA
the extracted work comes from enthalpy 
only, and the efficiency is given by (\ref{guppi-1}). 
Since the maximal work is reached for $\v(x_2=0.6714)>1$, Eq.~(\ref{hazare}) means that 
the kinetic energy increases due to
work-extraction.

The efficiency $\eta(x_2=0.6714)$ at the maximal work in Fig.~\ref{fig2.1} can be calculated 
from (\ref{maxim-2}, \ref{guppi-1}). For parameters of Fig.~\ref{fig2.1}, we get 
\BEA
\label{baktr}
\eta(x_2=0.6714)=0.4833.
\EEA
It is seen that the efficiency (\ref{baktr}) 
at the maximal work is smaller than the efficiency (\ref{pashto})
extracted from the kinetic energy only. 

\comment{
Let us now discuss the choice: $x_2^*=0.6331$, where $\v(x_2^*)=1$. This
choice will lead to the work-extraction from enthalpy at zero change of
the kinetic energy in (\ref{vedma-1}). We get from (\ref{vedma-1},
\ref{guppi-1}) [cf.~({\bf 16})]:
\BEA
\label{guppi-2}
\eta(x_2^*)=1-\a_2^{1-\g}(x_2^*).
\EEA
Since the work grows monotonously as Fig.~\ref{fig2.1} shows, 
$\eta(x_2^*)$ is a lower bound for the efficiency at the 
maximal work:
\BEA
\label{guppi-3}
\eta(x_2)>1-\a_2^{1-\g}(x_2^*).
\EEA
For parameters of Fig.~\ref{fig2.1}, we get $\eta(x_2^*)=1-(1 + 1.5 0.6331)^{-0.8}=0.4138$, which
is expectedly smaller than (\ref{baktr}).

For parameters of Fig.~\ref{fig2.1} it also holds that
\BEA
\label{guppi-4}
\eta(x_2)>1-\a_2^{1-\g}(x_2)=0.4273.
\EEA
But note that (\ref{guppi-3})
is based on the monotonic growth of work [cf.~Fig.~\ref{fig2.1}], 
but (\ref{guppi-4}) does not necessarily follow from (\ref{guppi-3}), since
$\a_2^{1-\g}(x_2)<\a_2^{1-\g}(x_2^*)$.
}

\begin{figure}[ht]
\includegraphics[width=7cm]{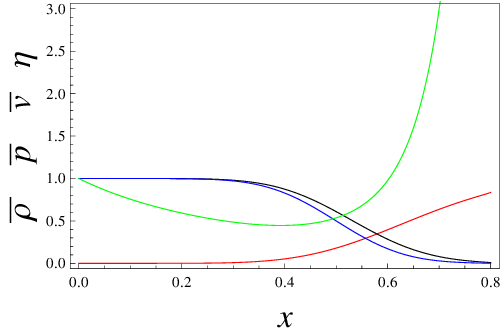}
\caption{ 
Dimensionless density $\r$ (black curve), pressure $\p$ (blue curve), 
velocity $\v$ (green curve), and the efficiency $\eta$ (red curve, defined by (\ref{guppi-1}))
versus $x$ obtained from solving {\bf (25, 26)} 
for $x\in [0,0.8011]$ under the same parameters as in Fig.~\ref{fig2.1},
but now $\ma=10^{-3}\times\frac{1}{7}$, i.e. the initial Mach number is very low.\\
The dimensionless velocity $\v(x)$ reaches the sonic
value {\bf (21)} at the end point $x=x_2=0.8011$, where the work attains its maximum {\bf (22)}. 
The density and pressure decay monotonously, i.e. the work-extraction always takes place also from
enthalpy. 
}
\label{fig3.1}
\end{figure}

We emphasize that the efficiency at the maximal work (\ref{maxim-2}) 
can be close to $1$. Indeed, recalling in (\ref{maxim-2})
that in the quasi-1d situation we have $\sigma=\v_{\rm tr}=0$
we get:
\BEA
\eta=1+\frac{(\g-1)\ma}{2}-\frac{(\g+1)}{2}\, \left(\ma\,\a_2^{-2}\right)^{\frac{\g-1}{\g+1}}.
\EEA
Hence $\eta\to 1$ for initially vanishing Mach number $\ma\to 0$.
Fig.~\ref{fig3.1} illustrates this situation: the maximal work {\bf
(22)} is reached at $x_2=0.8011$, where $\v(x_2=0.8011)>1$. The
efficiency at the maximal work equals $\eta(x_2=0.8011)=0.8377$; see
Fig.~\ref{fig3.1}. 

For parameters of Fig.~\ref{fig3.1} the density $\r(x)$ (and pressure
$\p(x)$) monotonously decay. Hence the scenario where work is extracted
from the kinetic energy only is absent: there is always a contribution
to work coming from enthalpy. Moreover, this is the main contribution into the
work.  Thus no comparison with the efficiency {\bf (19)} of the
work-extraction from kinetic energy can be carried out. 

It remains to stress that $\v_{\rm m}>1$ holds for many
reasonable values of parameters, but not always.  E.g. if in parameters
of Fig.~\ref{fig3.1} we increase the initial Mach number to
$\ma=\frac{95}{140}$ (which is still subsonic, but it already close to
the sonic threshold), the velocity (\ref{maxim-1}) at the maximal
work-extraction equals $\v_{\rm m}=0.9727\leq 1$. Since $0.9727 \approx 1$, 
this example illustrates the attainability of bound (\ref{ragor}).

\subsection{7. Sonic output velocity}

In {\bf Fig.~3} (of the main text) we saw that under a sufficiently strong
negative force $F(x)<0$ the output axial velocity $v(x)$ can reach the
sonic value.  This reachability is studied here in more detail.  Let us
integrate {\bf (27)} from $x_1$ to $x$, and write it using {\bf (25)}
\begin{gather}
\label{kora}
\frac{1}{\r^{2}(x)\a^{2}(x)}-1  +\frac{2 [\r^{\gamma-1}(x)-1]}{\ma(\gamma-1)}
={\cal F}(x), \\
{\cal F}(x)\equiv\frac{2}{\g\ma}\int_{x_1}^x \frac{\f(y)}{\r(y)}.
\end{gather}
Eq.~(\ref{kora}) is a quadratic equation for $\r^2(x)$ for $\gamma=3$, 
which is not close to the air value $\gamma=1.4$, but otherwise 
is physically sensible. It is solved as
\BEA
2\r^2(x)&=&1+\ma(1-{\cal F}(x))\nonumber \\
&\pm& \sqrt{\left[\, 1+ \ma(1-{\cal F}(x)) \,\right]^2-\frac{4\ma}{\a^2(x)} }.
\label{okun}
\EEA
The choice of signs in (\ref{okun}) is regulated by $\r(x_1)=1$, which
implies the + sign in (\ref{okun}) for initially subsonic velocities
$\ma<1$. If $\f(x)<0$ and $|\f(x)|$ is sufficiently large at least
for some $x$, then it is possible to nullify the square-root in
(\ref{okun}): 
\BEA
2\r^2(x)=1+\ma(1-{\cal F}(x))=2{\rm M}_1/\a(x).
\EEA
This expression is equivalent to the local speed of sound $v^2_{\rm
s}(x)=3p(x)/\rho(x)$, as seen using {\bf (25, 6)}.

\subsection{8. Implications of the Bernoulli equation for the incompressible situation}
\label{bern}

Here we shall work out some implications of the Bernoulli equation for
the incompressible situation. The equation applies to the flow shown in
Fig.~\ref{fig11} (after the singularity (\ref{barba}) of the force on
$x_0$ is removed). 

The incompressible situation reads:
\BEA
\rho={\rm const}.
\label{inca}
\EEA
Now any potential force can be written as
\BEA
\label{kosher2}
\vec{F}=
-\rho\vec{\nabla}U(x,y,z)=-\vec{\nabla}[\rho U(x,y,z)], 
\EEA
where $U(x,y,z)$ is the suitable potential. In particular, 
any force that depends only on $x$ can be written as in (\ref{kosher2}).
We shall assume that no potential is present initially, while its final
value is independent from $(y,z)$:
\BEA
\label{uu}
U(x_1,y,z)=0 \qquad U(x_2,y,z)=U_2.
\EEA
Conditions (\ref{uu}) are consistent with having localized force 
inside of the flow volume; see Fig.~\ref{fig11}.

Recall that under (\ref{inca}) the internal energy is constant and hence drops out from conservation laws \cite{landau-1}; only the term $\frac{p}{\rho}$ 
is relevant \cite{landau-1}. Also, the entropy is not involved. The potential $U(x,y,z)$ can be incorporated into the Bernoulli equation \cite{landau-1}:
\BEA
\label{bernoulli4}
\frac{\vec{v}^2(x,y,z)}{2}+\frac{p(x,y,z)}{\rho} +U(x,y,z)={\rm const}.
\EEA
Assuming that the flow lines are continuous and that for any point on $A(x_2)$ there is a unique point 
on $A(x_1)$ related by a flow line, we get from (\ref{bernoulli4}, \ref{uu}):
\begin{gather}
\frac{\vec{v}^2(x_2,y,z)}{2}+{p_2}\,\frac{\widetilde{p}(y,z)}{\rho}+U_2 = \frac{v_1^2}{2}+\frac{p_1}{\rho},
\label{fuchik2}
\end{gather}
where we recall 2 definitions:
\begin{gather}
\label{oror}
p(x_1,y,z)=p_1, \qquad p(x_2,y,z)=p_2\widetilde{p}(y,z),\\
\langle\widetilde{p}\rangle
\equiv \int_{{A}_2} \frac{{\rm d}y\,{\rm d}z}{a_2}\,\widetilde{p}(y,z)=1.
\end{gather}

The energy conservation law reads [see (\ref{ursula})]:
\BEA  
\label{eq:02b}
\frac{v^2_{1}}{2}+\frac{p_{1}}{\rho}
  =\frac{v^2_{2}+\hat{v}^2}{2}+\frac{p_{2}}{\rho}-\frac{\int {\rm d}V\, \vec{v}\cdotp\vec{F}  }{a_{1}\rho v_{1}},\\
\hat{v}^2 \equiv \int_{{A}_2}\frac{{\rm d}y\,{\rm d}z}{a_2}\, [{v}_{y}^{2}(x_2,y,z) +{v}_{z}^{2}(x_2,y,z)] .
\EEA
Denoting
\begin{eqnarray}
  \label{eq:42}
\vec{v}_{\rm tr}(y,z)\equiv (0,v_{y}(x_2,y,z),v_{z}(x_2,y,z)),
\end{eqnarray}
we conclude from (\ref{eq:02b}, \ref{fuchik2}):
\begin{gather}
\frac{\vec v^2_{\rm tr}(y,z)-\hat{v}^2}{2}+\frac{p_{2}}{\rho} 
  \left( {\widetilde{p}(y,z)}-1\right)
=-U_2-\frac{\int {\rm d}V\, \vec{v}\cdotp\vec{F}  }{a_{1}\rho v_{1}},
  \label{eq:533}
\end{gather}
due to $v_{x}(x_{1},y,z)=v_{y}(x_{1},y,z)=0$ and (\ref{oror}).
Integrating (\ref{eq:533}) by $\int_{A_2}\frac{{\rm d}y\, {\rm d} z}{a_2}$ we conclude that 
\BEA
\label{dubac2}
U_2+\frac{\int {\rm d}V\, \vec{v}\cdotp\vec{F}  }{a_{1}\rho v_{1}}=0,\\
\label{bac2}
\frac{\vec v^2_{\rm tr}(y,z)-\hat{v}^2}{2}+\,\frac{p_{2}}{\rho_{2}} 
  \left( {\widetilde{p}(y,z)}-1\right)=0.
\EEA
Eq.~(\ref{bac2}) shows that homogeneous pressure $\widetilde{p}(y,z)=1$ leads to zero transverse velocities:
$\vec v^2_{\rm tr}(y,z)$ not depending on $(y,z)$ means $\vec v^2_{\rm tr}(y,z)=0$, since $\vec v^2_{\rm tr}(y,z)$
has to nullify on the boundaries of $A(x_2)$. 

Note that (\ref{dubac2}) automatically holds within the quasi-1d approach, where 
\begin{gather}
\frac{\int {\rm d}V\, \vec{v}\cdotp\vec{F}  }{a_{1}\rho v_{1}}
=\frac{     \int_{x_1}^{x_2}\d x\, a(x)v(x)F(x)   }{a_{1}\rho v_{1}}
=\frac{1}{\rho }{\int_{x_1}^{x_2}\d x\, F(x)   }.
\end{gather}

\begin{figure}[ht]
\includegraphics[width=7cm]{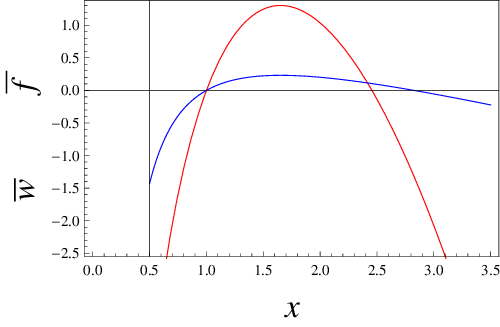}
\caption{ 
Dimensionless work $\w$ (red curve) and force $\bar{f}$ (blue curve)
versus $x$ for $\ma=0.3$. Here $\w$ and $\bar{f}$ are obtained from (resp.) 
(\ref{vedma10}) and (\ref{karasultan}) under $\sigma=\v_{\rm tr}=0$.
}
\label{fig6.1}
\end{figure} 

\subsection{9. Work-extraction in a cylindrical tube: relations with d'Alembert paradox}

\subsubsection{9.1 No work-extraction from kinetic energy}

Recall that the choice of the control volume $B$ in {\bf Fig.~1} is
conventional (i.e. other choices are also possible) and is subject to 2
conditions discussed after {\bf (1)}. Here we choose the control volume
differently. We take it so large as it includes the domain $\Omega$ of
work-extraction and it is cylindrical, i.e.  the cross-section $A(x)$
and its area $a(x)$ are constants. Since $B$ is defined along the flow
lines, this choice of a large $B$ will first of all violate assumption {\bf 5}
[see {\bf (2)}], since now the output velocity $v_x(x_2,y,z)$ will
essentially depend on $(y,z)$: for $(y,z)$ close to the boundary of
$A(x_2)$ the flow will be practically unperturbed, $v_x(x_2,y,z)\simeq
v_x(x_1)$ but closer to the center of $A(x_2)$ we do expect serious
differences between $v_x(x_2,y,z)$ and $v_x(x_1)$. 

However, for methodological reasons it is still interesting to assume a
cylindrical shape of $B$ and implement all assumptions including {\bf
5}. To avoid the above inconsistency with assumption {\bf 5}, $B$ can be
regarded as a real cylindrical tube in which the fluid flows in the
stationary regime. Now (\ref{eq:03a}) is useful, since $\zeta=0$ due to
$A_3=0$. Once $B$ is a cylinder, we have $\a_2=a_2/a_1=1$, and the
conservations of mass, entropy and momentum read in the dimensionless
form [cf.~{\bf (14, 15)}]:
\BEA
  \label{sultan}
&& \r\v = 1, \qquad \p_2=\r_2^{\g}\, e^{\sigma }, \\
&& \g \ma (1-\v_2)+1-\p_2=\bar{f}, \qquad \bar{f}\equiv
\frac{-\int{\rm d}V\, {F}_x}{a_1p_1},~~~~
  \label{nursultan}
\EEA
while the energy conservation leads to the definition of work {\bf (16)} that we copy below:
\BEA
\label{vedma10}
\w=
1-\v_2^2-\v_{\rm tr}^2+\frac{2}{\ma(\g-1)}(1-e^{\sigma}\v_2^{1-\g}).
\EEA
Recalling that $\sigma>0$ and $\gamma>1$ (which have a thermodynamic
origin) and the subsonic condition $\ma<1$, we see from (\ref{vedma10})
that no work-extraction from kinetic energy is possible, since $\w<0$
for all $\v_2<1$. This fact is easily seen from differentiating
(\ref{vedma10}) over $\v_2$. 

\subsubsection{9.2 Relations with d'Alembert paradox}

The above conclusion relates to d'Alembert's paradox \cite{sedov-1}.
Recall the set-up of this paradox \cite{sedov-1}. One considers a
smooth body immersed into a cylindric tube and by-passed by a
dissipationless fluid. Formally, no volume force is present here, but the
effective force appears due to integration of the momentum conservation
relation over the volume of the body that is excluded from the control 
volume (i.e. the cylindric tube). Due to
boundary conditions no contribution from the body enters into the energy
equation. Hence $\w=0$. One also assumes that both input and output flows are
homogeneous, i.e. $\sigma=\v_{\rm tr}=0$. Then $\w=0$ from
(\ref{vedma10}) leads to $\v_2=1$, which together with (\ref{sultan})
implies from (\ref{nursultan}): $\int{\rm d}V\, {F}_x=0$, i.e. the
$x$-component of the force acting on the body nullifies \cite{sedov-1}. 

Note that generally $\int{\rm d}V\, {F}_y\not =0$ \cite{sedov-1}. This is
seen from (\ref{ushi}) even if we put there
$\left\langle\,\widetilde{\rho}(y,z)\,v_y(x_2,y,z)\,\right \rangle=0$
assuming a homogeneous output. 

\subsubsection{9.3 Work-extraction from enthalpy}

Let us now return to (\ref{sultan}--\ref{vedma10}) and continue to assume there that 
$\sigma=\v_{\rm tr}=0$. Then we get from (\ref{nursultan}):
\BEA
\label{karasultan}
\g \ma (1-\v_2)+1-\v_2^{-\g}=\bar{f},
\EEA
i.e. given the external force (given $\bar{f}$) we can determine $\v_2$ from 
(\ref{karasultan}) and find out the work $\w$. Fig.~\ref{fig6.1} shows that
for a given $\ma<1$ there are forces $\bar{f}>0$ that can lead to work-extraction $\w>0$ under $\v_2>1$.
Thus the only possible scenario here is the work-extraction from enthalpy with increasing kinetic energy.


\begin{thebibliography}{99}
\bibitem{zhuk}
Indeed, $\langle \widetilde{\rho}\,
  \ln\frac{\widetilde{\rho}}{\widetilde{p}} \rangle\geq 0$, because it is
a relative entropy between two distributions. To show 
$L\equiv \langle\widetilde{\rho}\,\ln\widetilde{\rho}\rangle\geq 0$
from (\ref{avo}), minimize $L$
over $\widetilde{\rho}$ as independent functional variable keeping track of
(\ref{avo}) via a Lagrange
multiplier: $L$ is minimized for $\widetilde{\rho}=1$, and $L=0$ at
the minimum.

\comment{discretize (\ref{avo}) writing
it as $\sum_{k=1}^N \epsilon_k \widetilde{\rho}_k=1$, where
$\sum_{k=1}^N \epsilon_k=1$, $\epsilon_k\geq 0$ and
$\widetilde{\rho}_k\geq 0$.  Now $L=\langle\widetilde{\rho}\,\ln\widetilde{\rho}\rangle=\sum_{k=1}^N \epsilon_k
\widetilde{\rho}\ln\widetilde{\rho}_k\geq 0$ follows after minimizing $L$
over $\widetilde{\rho}_k$ as independent variables keeping track of
$\sum_{k=1}^N \epsilon_k \widetilde{\rho}_k=1$ via a Lagrange
multiplier: $L$ is minimized for $\widetilde{\rho}_k=1$, and $L=0$ at
the minimum.}
\end{thebibliography}
\begin{thebibliography}{99}
\bibitem{choban}  Not to mention that the force localization in (\ref{barba}) is in potential
conflict with incompressibility (\ref{rhorho}). We shall not expand on drawback further, since
the set-up does have more straightforward problems.
\bibitem{drud}Drude's relation (\ref{drude}) can be rigorosuly deduced from classical mechanics of a point particle.
While this derivation is interesting by itself and hence is reproduced below, it by no means implies that this relation
also holds in hydrodynamics. Note that ``particles" of a compressible
ideal fluid are not independent mechanical points, rather they are fluid
parcels that move with conserving their mass (but not the shape), and
interact via their thermally isolated (adiabatic) boundaries. \\
A point particle with mass $m$ that moves in the positive $x$-direction
and is subject to a localized (delta-function) force $-f\delta(x)$
located at $x=0$ and directed against the motion: $f>0$. This force
may be generated by a step-function potential $U(x)=f\vartheta
(x)$. Now energy and momentum conservation imply, respectively:
\begin{eqnarray}
  \label{eq:43}
  \frac{m}{2}(v_1^2-v_2^2)=f\equiv w, \qquad 
m(v_1-v_2)=f/v(0),
\end{eqnarray}
where $w$ is the work, while $v_1=v(x<0)$ and $v_1=v(x>0)$ are,
respectively, the initial and final velocities. The second 
relation in (\ref{eq:43}) is got after integrating the Newton
equation of motion in the vicinity of $x=0$.
Two equations in (\ref{eq:43}) imply Drude's relation 
$v(0)=\frac{v_1+v_2}{2}$ for
the velocity at the work-extraction point; hence $v(x)$ is continuous
in the vicinity of $x=0$, though changes there fastly. It is seen that
the conservaion laws allow the extraction of the full kinetic energy,
i.e. $v_2\approx 0$ and $w\approx \frac{mv_1^2}{2}$, provided that 
${mv_1^2/2}\lesssim f $.
\end{thebibliography}
\end{document}